\documentclass[aps,preprint,floatfix,nofootinbib,showpacs]{revtex4-1}
\pdfoutput=1
\usepackage{graphicx,color,rotating}
\usepackage{hyperref}
\usepackage{epsfig}

\begin{document}
\renewcommand{\thefootnote}{\arabic{footnote}}

\title{Dark Matter with multi-annihilation channels and 
AMS-02 positron excess and antiproton }
\author{
Yu-Heng Chen$^1$, Kingman Cheung$^{1,2,3}$, and Po-Yan Tseng$^1$}
\affiliation{
$^1$ Department of Physics, National Tsing Hua University,
Hsinchu 300, Taiwan \\
$^2$ Division of Quantum Phases and Devices, School of Physics,
Konkuk University, Seoul 143-701, Republic of Korea \\
$^3$ Physics Division, National Center for Theoretical Sciences,
Hsinchu, Taiwan
}
\date{\today}

\begin{abstract}
AMS-02 provided the unprecedented statistics in the measurement of the
positron fraction from cosmic rays. That may offer a unique
opportunity to distinguish the positron spectrum coming from various
dark matter (DM) annihilation channels, if DM is the source of this
positron excess. Therefore, we consider the scenario that the DM can
annihilate into leptonic, quark, and massive gauge boson channels
simultaneously with floating branching ratios to test this hypothesis.
We also study the impacts from MAX, MED, MIN, and DC diffusion models as
well as from isothermal, NFW, and Einasto DM density profiles on our
results. 
We found two parameter regions that can satisfy both AMS-02 
$\frac{e^+}{e^++e^-}$ and $\bar{p}/p$ datasets at  95\% CL.
i) Under the NFW-MIN combination with $M_{\chi}\subset[10,30]$ TeV.
ii) Under the Einasto-DC combination with $M_{\chi}\subset[500,1500]$ GeV.
\end{abstract}

\maketitle

\section{Introduction}

Up to now the evidences of the existence of dark matter (DM)
only come from the gravitational effects. In the early 1950s,
people noticed the 
anomaly in the rotational velocity in the Milky Way and other 
spiral galaxies, and one logical reason is that 
there are non-luminous matters suffusing our
universe. Currently the Planck satellite \cite{planck} pinned down 
that 26.8\% of our universe is made up of DM. The Bullet cluster and N-body
simulations may reveal some self-interaction of DM particles.

From the particle physicists' point of view, the DM particle might have
interactions with the standard model (SM) particles, that will 
lead to interesting phenomenology. 
The DM direct detection experiments, like XENON100\cite{xenon100},
CDMS II\cite{cdms} etc., can probe the interactions between the DM and
nucleon. Searching for the DM signal as missing energies have been performed at collider experiments
such as ATLAS \cite{atlas_1,atlas_2} and CMS \cite{cms}. In this work,
we focus on DM indirect detection. Since the DM suffuses our
galaxy, they probably annihilate with each other into SM
particles. 
In this case, the antimatters, like positrons or antiprotons,
would be the ideal signals for searching for 
these processes, because the generation of
antimatters is rare from ordinary astrophysical processes. The Alpha
Magnetic Spectrometer (AMS-02) is a particle detector, which was
designed to search for the antimatter and determine the composition and
flux of the cosmic ray. The AMS-02 published the measurements
of electron and positron results with unprecedented statistics in 2014. 
Their results confirmed the 
positron excess with energy higher than 10 GeV up to 500 GeV, which
was first indicated by the PAMELA on 2008~\cite{pamela,pamela2008}. 
On the contrary, based on PAMELA and most recent AMS-02~\cite{ams02_antip} 
observation there is no significant excess of antiproton in 
the cosmic rays from the prediction of
conventional astrophysical processes.

The DM annihilation is one of the viable interpretations for the observed 
positron excess,
however it depends on which kinds of the SM particles that 
the DM particles annihilate into.
The antiproton signal in general co-exists with the positron signal 
as parts of the annihilation products, that is severely constrained 
by the exisiting 
antiproton observations. This is one of the main difficulties in
explaining the positron excess by DM annihilation.
Furthermore, the newly released $\frac{e^+}{e^++e^-}$ data from AMS-02 has 
so high statistics that it may be used to distinguish various DM
annihilation channels. In this work, we set up a
model-independent DM annihilation framework, in which the DM could
annihilate into lepton channels ($e^+e^-$, $\mu^+,\mu^-$,
$\tau^+\tau^-$), quark channels ($t\bar{t}$, $b\bar{b}$), and massive
gauge boson channels ($W^+W^-$, $ZZ$) for a single annihilation with
varying branching ratios (BR). Through fitting to the $\frac{e^+}{e^++e^-}$ and
$\bar{p}/p$ data, we investigate if this DM scenario
can survive, if it does, which one is the preferred channel.
Here we use the $b\bar{b}$ channel to represents other lighter quark channels, because with DM mass above few hundreds GeV, the positron and anitproton spectra from $b\bar{b}$ channel are similar to those spectra from other lighter quark channels.

Note that AMS-02 very recently released the preliminary data on
the ratio $\bar p/p$ \cite{ams02_antip}, which also seems to have 
an unexpected rise at the first glance. However, it was shown to be
consistent with the astrophysical background and conventional 
diffusion model within uncertainties \cite{ams02_antip_1,ams02_antip_2,ams02_antip_3}.
We will include the most recent AMS-02 $\bar p/p$ data in our analysis.

This work is organized as follows. We briefly describe the AMS-02
$\frac{e^+}{e^++e^-}$ data and background models in the next section
and DM model in Sec.III. The fitting results of $\frac{e^+}{e^++e^-}$
data shall be shown in Sec.IV. In Sec.V, we start considering the 
constraint from AMS-02 $\bar{p}/p$ data. In Sec.VI, we discuss the
influences from density profiles. Then we conclude in Sec.VII.

\section{AMS-02 positron data and background}
In September 2014, the AMS collaboration published their high
statistics measurement of the $\frac{e^+}{e^++e^-}$ in primary cosmic
rays of 0.5-500 GeV~\cite{ams02_frac}. The $\frac{e^+}{e^++e^-}$
increases and exceeds the astrophysical cosmic ray expectation for 
energy larger than about 10 GeV. This trait is consistent with the
earlier observations by PAMELA~\cite{pamela,pamela2008} and 
Fermi-LAT\cite{fermi_lat}. Furthermore, the AMS-02 provides 
the most precise measurement so far such that it offers more 
information and tighter constraints on the origin of positron excess.

The uncertainty of the high energy cosmic positron flux, which is
originated from astrophysical sources and presumingly generated from
the collision of high energy proton and interstellar medium
(here we denote as the background), is still
large. It is due to lack of knowledge of the sources and propagation
models for the positron flux. 
%
%
The proton flux has been precisely measured by many experiments in the
past and the most recently by AMS-02. The secondary electron and
positron flux could be calculated straight-forwardly from the interactions of
proton with interstellar material (mostly hydrogen and
helium). However, there are large uncertainties 
in the propagation of the secondary particles from the origin of production
to the top of atmosphere (TOA). 
Difference diffusion models will give different
positron spectra and as well as astrophysical background of
$\frac{e^+}{e^++e^-}$ when we deal with the AMS-02 positron fraction
data.

\begin{figure}[th!]
\centering
\includegraphics[width=5.4in]{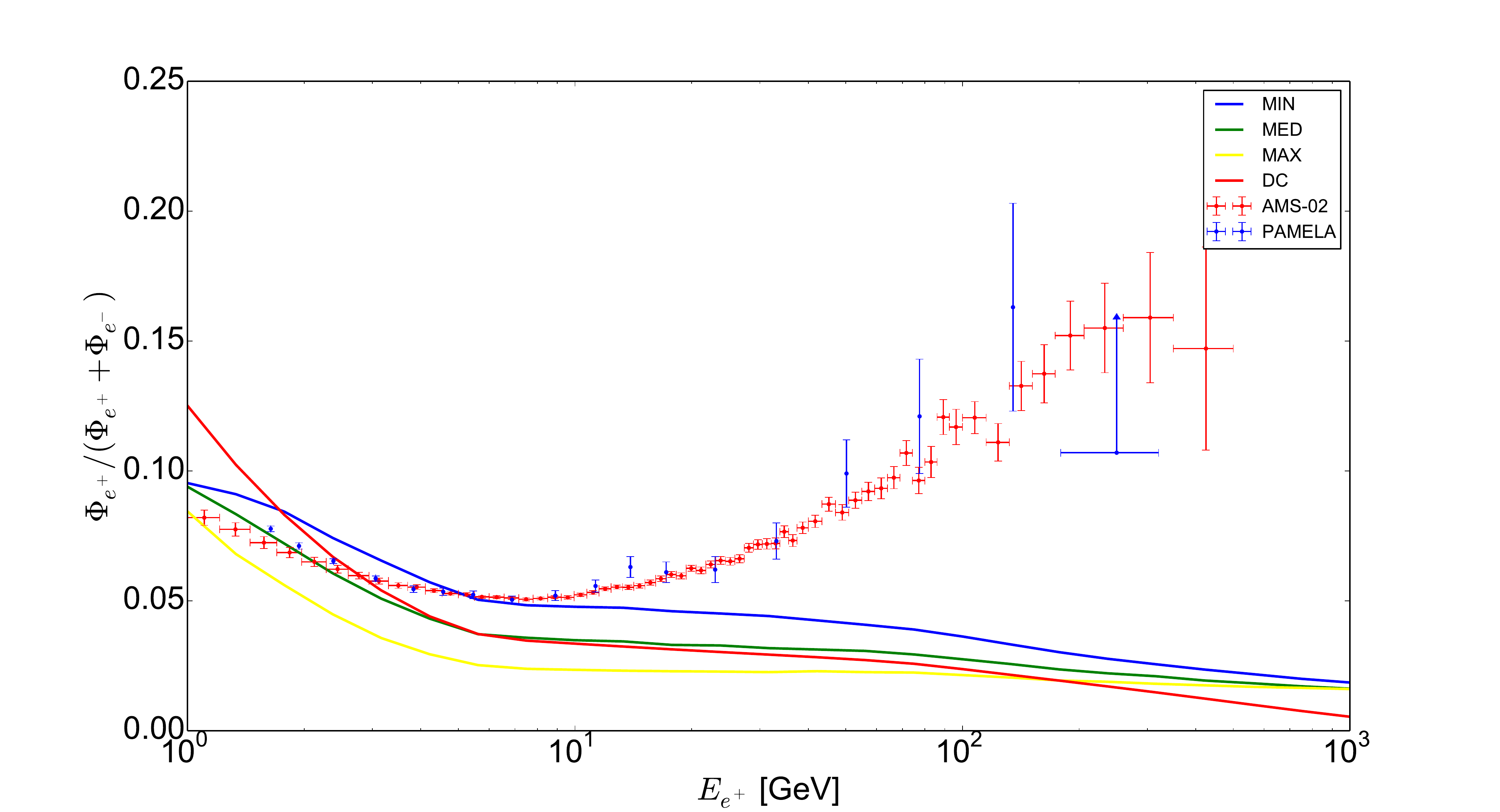}
\caption{ \small \label{bkgd} 
The astrophysical backgrounds for the observable of
$\Phi_{e^+}/(\Phi_{e^-}+\Phi_{e^+})$ using different diffusion models, 
corresponding to the MIN, MED, MAX, and DC diffusion models. 
Red crosses are the AMS-02 $\frac{e^+}{e^-+e^+}$
data~\cite{ams02_frac} and blue crosses are the PAMELA
$\frac{e^+}{e^-+e^+}$ data~\cite{pamela,pamela2008}.  
}
\end{figure}

In this work, we will consider four different diffusion models, namely
MIN, MED, MAX~\cite{diffusion,med,max_min}, and a diffusion convection
model (DC)~\cite{dc}. Each diffusion model is paired with a
background, which is obtained by using the method mentioned above and
computed through the Galprop package~\cite{galprop1,galprop2}, and
these backgrounds were shown in Fig.\ref{bkgd}. More description and
the parameters for these models will be given in the next section. 
In Fig.~\ref{bkgd}, we compare the four backgrounds of
$\frac{e^+}{e^++e^-}$ from the four diffusion models. The ordering of
background values of $\frac{e^+}{e^++e^-}$ is MIN $>$ MED $>$ MAX,
when the energy $E_{e^+}\geq$ 10 GeV. For the DC case, the background
is close to that of MED case but has a slight deficit, when
$E_{e^+}\geq$ 100 GeV.

The backgrounds of $\frac{e^+}{e^++e^-}$ from MIN, MED, and MAX
diffusion models are quite different from one another, in such a way that
we anticipate that they are able to cover or represent larger portions of
diffusion models in the market. 
Also, they are able to test whether
our results are diffusion-model independent. The reason we adopt
the DC diffusion model is that it utilized the more present B/C
data to constrain the cosmic-ray propagation parameters.

\section{Positron source from DM Annihilation}

The AMS-02~\cite{ams02_frac} provides a very precise measurement of the
positron fraction, such that it could discriminate different DM
annihilation channels and determine which one is preferred or ruled out.  
In order to do so, we consider the multi-annihilation 
channels DM scenario by assuming that 
the dark matter (DM) particles $\chi$ could annihilate into seven
standard model (SM) channels $t\bar{t}$, $b\bar{b}$, $W^+W^-$, $ZZ$,
$e^+e^-$, $\mu^+\mu^-$, and $\tau^+\tau^-$ from a single annihilation
process. The branching ratio for each channel can float between $0$ and
$1$, as long as they satisfy the sum rule
\begin{equation}
\label{sum}
\sum_{f=t,b,W,Z,e,\mu,\tau}\textrm{BR}(f\bar{f})=1 \,.
\end{equation}
The relation between the total DM annihilation cross section and 
each channel is specified as
$$
\left\langle \sigma v \right\rangle_{\chi \bar{\chi} \rightarrow f \bar{f}} =  
\textrm{BR}(f\bar{f}) \times \left\langle \sigma v \right\rangle_{tot} \,.
$$ 
These SM particles in the final state will further decay or
hadronize into positrons, antiprotons, or gamma-rays, 
 which will be added on to the spectra obtained by the 
conventional high energy cosmic rays.
Due to the distinctive spectral shape of these DM annihilation
products, we may be able to separate these DM signals from the
astrophysical cosmic rays and interpret as an indirect evidence of DM.
In general, the leptonic channels $e^+e^-$, $\mu^+\mu^-$, and 
$\tau^+\tau^-$ produce a relatively harder positron energy spectrum than 
the quark channels $t\bar{t}$ and $b\bar{b}$. The positron spectra 
from massive vector boson channels $W^+W^-$ and $ZZ$ sit
between the leptonic and quark channels.

We use the interface for model-independent analysis from
micrOMEGAs~\citep{micromegas}, where the Pythia~\citep{pythia} was
embedded to generate the positron and antiproton spectra of each DM
annihilation channel. Several DM density profiles will be considered
in this work, including NFW~\cite{nfw}, isothermal~\cite{iso}, and
Einasto~\cite{ein1,ein2}. Nevertheless, most of the time we focus on the
NFW profile. The NFW and isothermal can be parametrized by
$$
\rho(r)\,=\,\rho_{\odot}\left\lbrace\frac{r_{\odot}}{r} \right\rbrace^{\gamma}
\left\lbrace \frac{1+(r_{\odot}/r_s)^{\alpha}}{1+(r/r_s)^{\alpha}} 
\right\rbrace^{(\beta-\gamma)/\alpha}  \,,
$$
where $\rho_{\odot}=0.3\,\rm{[GeV/cm^3]}$ is the DM density at the Sun, 
$r_{\odot}=8.5\,\rm{[kpc]}$ is the distance from the Sun to the Galactic Center
(GC).
The NFW (isothermal) density profile can be obtained by choosing the 
parameters $\alpha=1, \beta=3, \gamma=1$, and $r_s=20$[kpc] 
($\alpha=2, \beta=2, \gamma=0$, and $r_s=3.5$[kpc]). 
On the other hand, the expression for the Einasto density profile is
$$
\rho(r)\,=\,\rho_s e^{\left\lbrace -\frac{2}{\alpha} 
\left[ \left( \frac{r}{r_s} \right)^{\alpha}-1  \right]  \right\rbrace }\, ,
$$
with $\rho_s=0.3\,\rm{[GeV/cm^3]}$, $r_s=8.5\,\rm[kpc]$, and $\alpha=0.17$,
that will give $\rho(r=r_{\odot})=\rho_{\odot}$.

In micrOMEGAs~\cite{micromegas}, the energy spectrum of the positrons 
is obtained by solving the diffusion-loss equation while keeping only 
the two dominant contributions: spatial diffusion and energy losses,
\begin{equation}
\label{diffusion}
-\nabla \cdot \left(K(E) \nabla \psi_{e^+} \right) 
-\frac{\partial}{\partial E} \left(b(E)\psi_{e^+} \right) \,=\,Q_{e^+}(\textbf{x}, E)\,,
\end{equation}
and
$$
K(E)\,=\,K_0 \beta(E)(\tilde{R}/1 \rm{GeV})^{\delta}\,,
$$
where $\beta$ and $\tilde{R}=p/q$ are the velocity and rigidity of the 
particle, respectively. The positron energy loss rate $b(E)$
is dominated by the synchrotron radiation in the galactic magnetic
field and by inverse Compton scattering on stellar and CMB photons. The
diffusion zone of cosmic rays is represented by a cylinder with
thickness $2L$ and radius $R$ (here we use $R=20\,
\rm{[kpc]}$)~\cite{diffusion}. The term $Q_{e^+}$ on the right hand side of
Eq.~\ref{diffusion} is the positron source term. 

The three diffusion models (MAX, MED, and MIN)~\cite{diffusion,med,max_min} 
that we are going to consider in this work, correspond to the 
diffusion parameters:
\begin{eqnarray}
{\rm MAX:}\,&& {\rm L}=15 \,, {\rm [kpc]}\,, K_0=0.0765\,, {\rm[kpc^2/Myr]}\,,
 \delta=0.46\,,\nonumber\\
{\rm MED:}\,&& {\rm L}=4 \,, {\rm [kpc]}\,, K_0=0.0112\,, {\rm[kpc^2/Myr]}\,,
 \delta=0.70\,,\nonumber\\
{\rm MIN:}\,&& {\rm L}=1 \,, {\rm [kpc]}\,, K_0=0.0016\,, {\rm[kpc^2/Myr]}\,,
 \delta=0.85\,,\nonumber
\end{eqnarray}
all of them satisfy the B/C observations.
Since the AMS-02 also provided the measurement of the B/C ratio, for
more comprehensive discussion of the influence of diffusion models,
we add one more diffusion model, the DC model from Ref.\cite{dc},
where the diffusion parameters were determined by fitting to the most
recent AMS-02 B/C data.

Under the DM with the multi-annihilation-channel scenario, it will 
contribute to the source term $Q_{e^+}$ in Eq.~(\ref{diffusion})
through the expression
\begin{equation}
\label{source}
Q^{DM}_{e^+}\;=\;\eta \left( \frac{\rho_{DM}}{M_{\chi}} \right)^2 
\left\langle \sigma v \right\rangle_{tot} \sum_{f=t,b,W,Z,e,\mu,\tau} 
\textrm{BR}(f\bar{f})\; \frac{dN^f_{e^+}}{dE_{e^+}}
\end{equation}
where $dN^f_{e^+}/dE_{e^+}$ is the positron energy spectrum from 
each annihilation event $\chi \bar{\chi}\rightarrow f \bar{f}$, 
and $\eta\;=\;1/2\;(1/4)$ for (non-)identical initial state. 
The $\rho_{DM}$ is the DM density profile, and $M_{\chi}$ is the DM mass.

\section{AMS-02 Positron fraction data and Fitting Result}
\label{III}

Here we attempt to use the DM scenario with multi-annihilation channels,
each of which has a varying branching ratio subject to the sum rule 
in Eq.~(\ref{sum}), to fit the
AMS-02 $\frac{e^+}{e^++e^-}$ data~\cite{ams02_frac}. The
varying parameters are the DM mass, the annihilation cross section, and the
branching ratios of seven annihilation channels: $M_{\chi}$,
$\left\langle \sigma v \right\rangle_{tot} $, and
$\textrm{BR}(f\bar{f})$. However, the branching ratios should satisfy the
sum rule in Eq.~(\ref{sum}), therefore, overall we have 8 free parameters.

First, we fix the DM density to the NFW profile and use 
  various diffusion models (MIN, MED, MAX, and DC) to compute the
  positron flux at TOA from DM annihilation. The positron source term
of DM annihilation channels can be described by
Eq.~(\ref{source}). Then we feed the source term into the diffusion
equation (\ref{diffusion}), and it is solved numerically by the
micrOMEGAs. Finally, we stack the DM-produced electron and
  positron flux on top of each diffusion-corresponding background flux
  to obtain the values of $\frac{e^+}{e^++e^-}$. Then we are able to
  compare with the AMS-02 positron fraction data.

\begin{figure}[th!]
\centering
\includegraphics[width=3.2in]{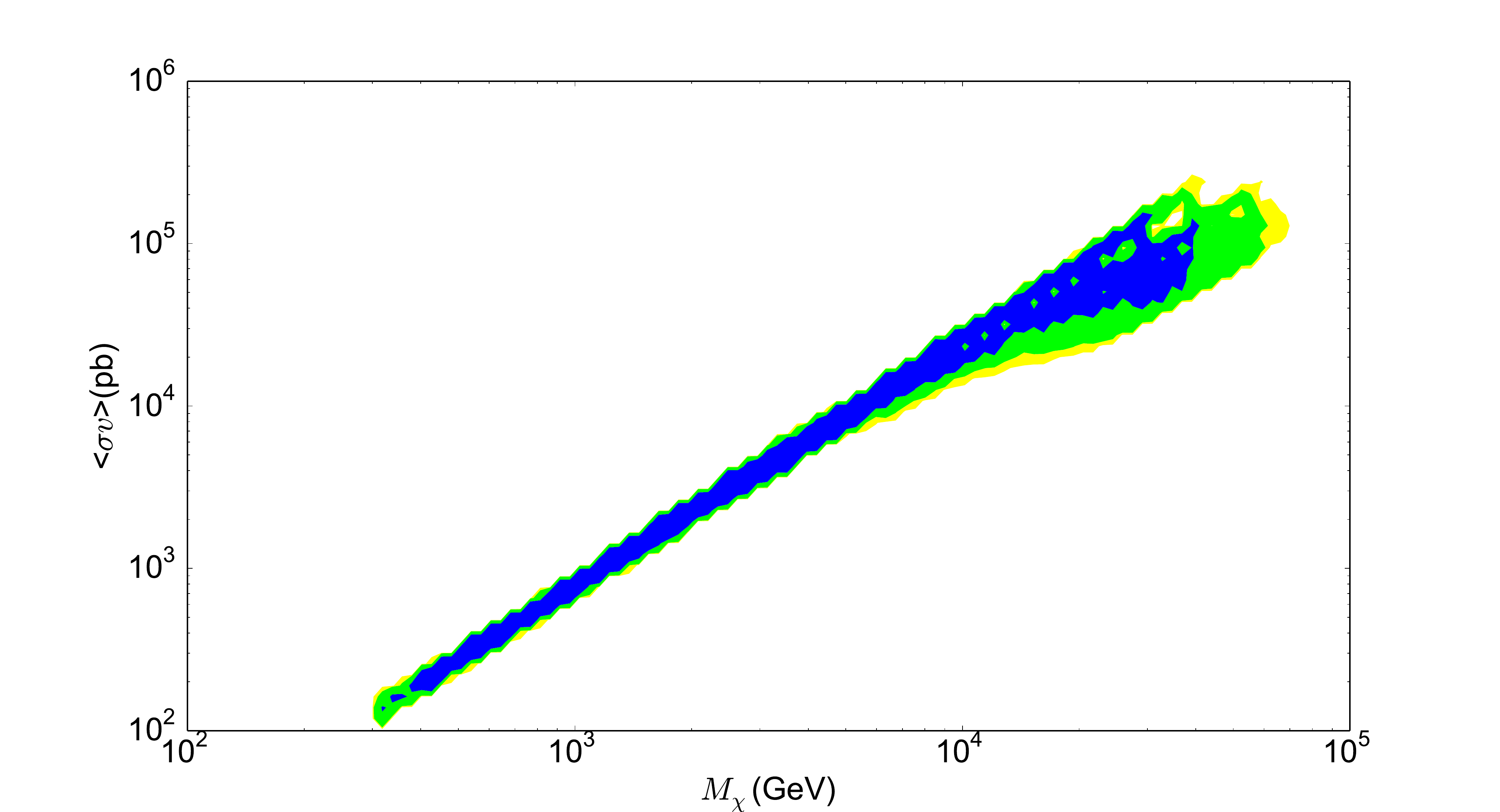}
\includegraphics[width=3.2in]{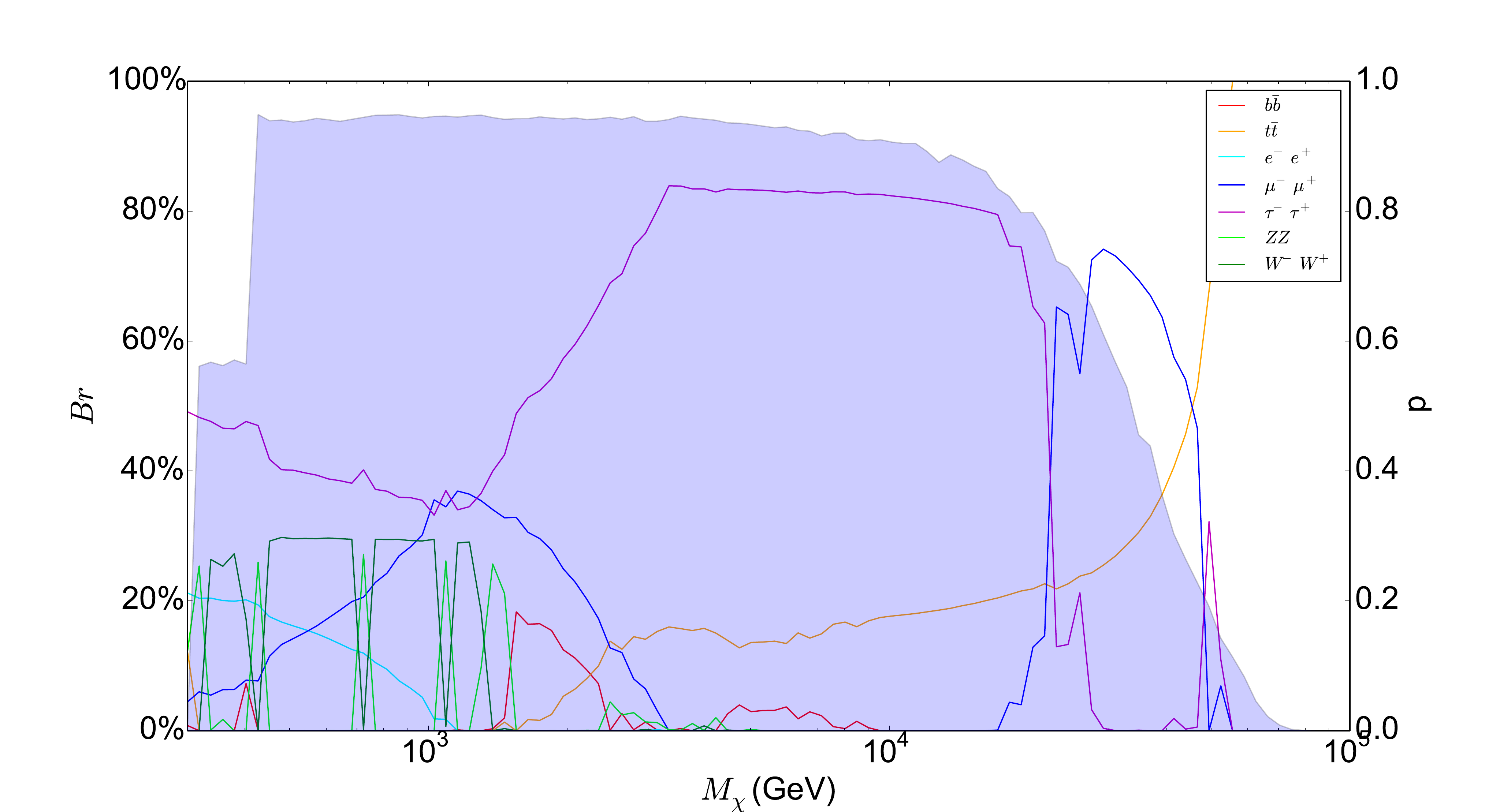}
\includegraphics[width=3.2in]{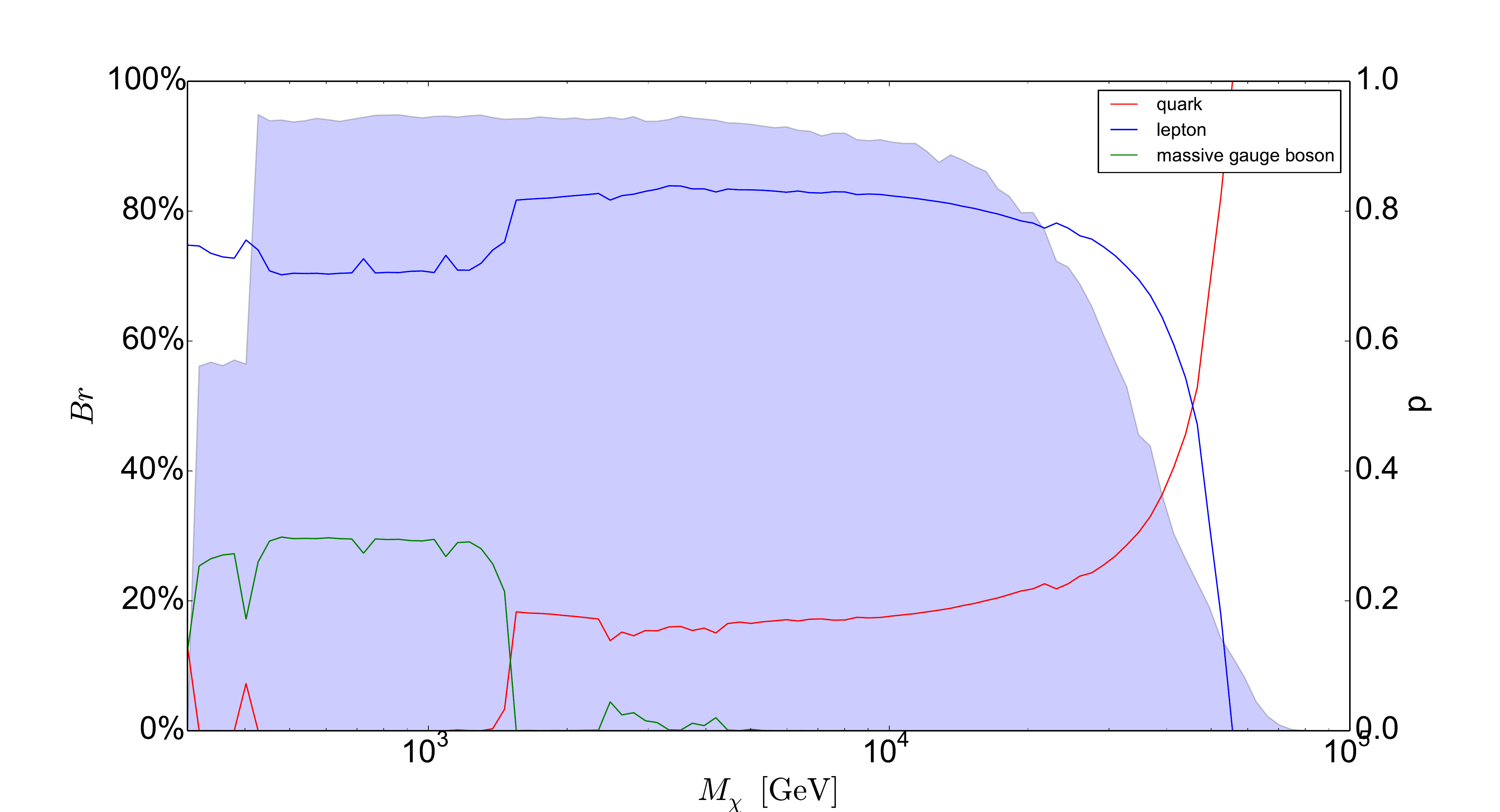}
\includegraphics[width=3.2in]{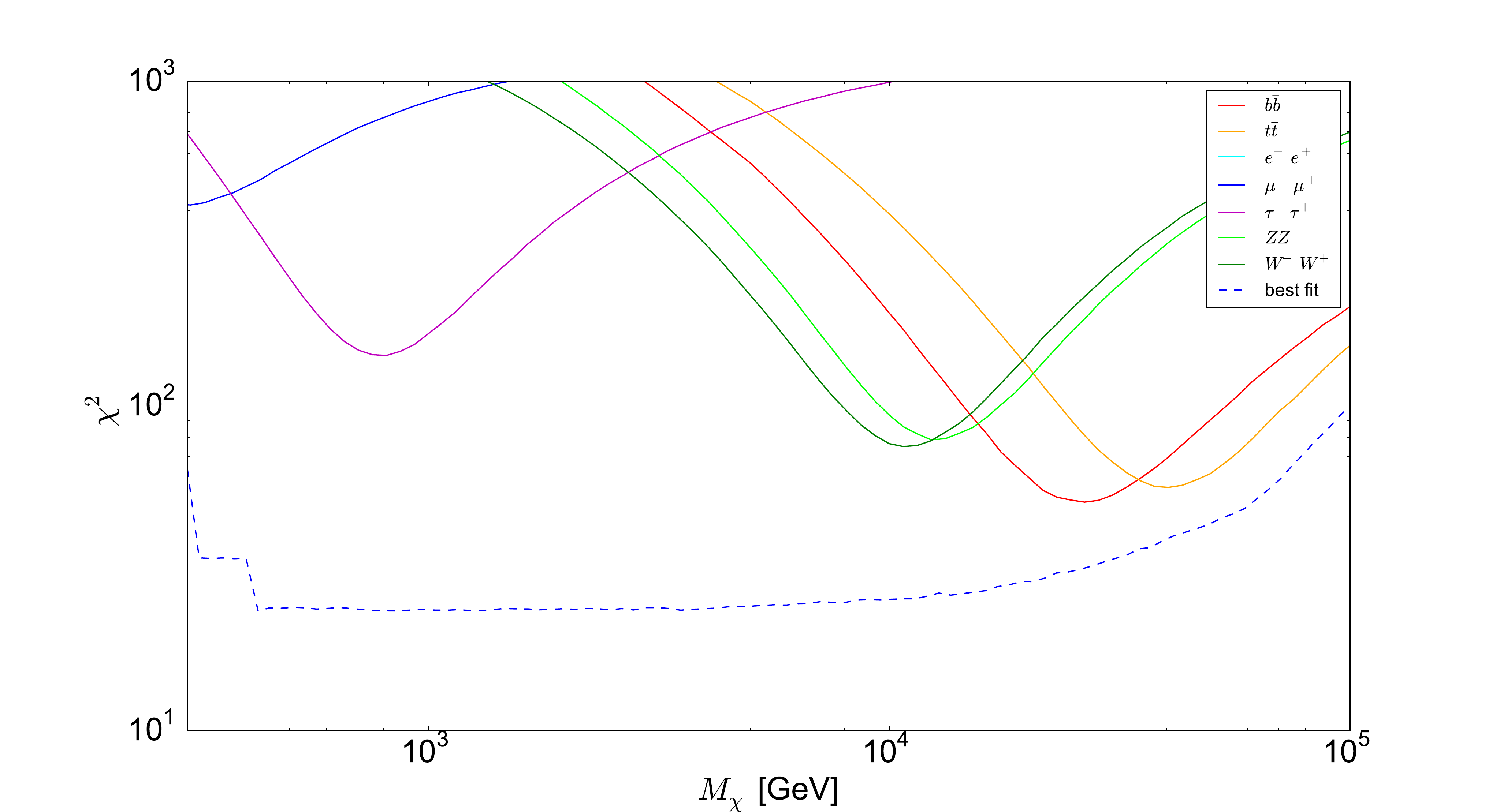}
\includegraphics[width=3.2in]{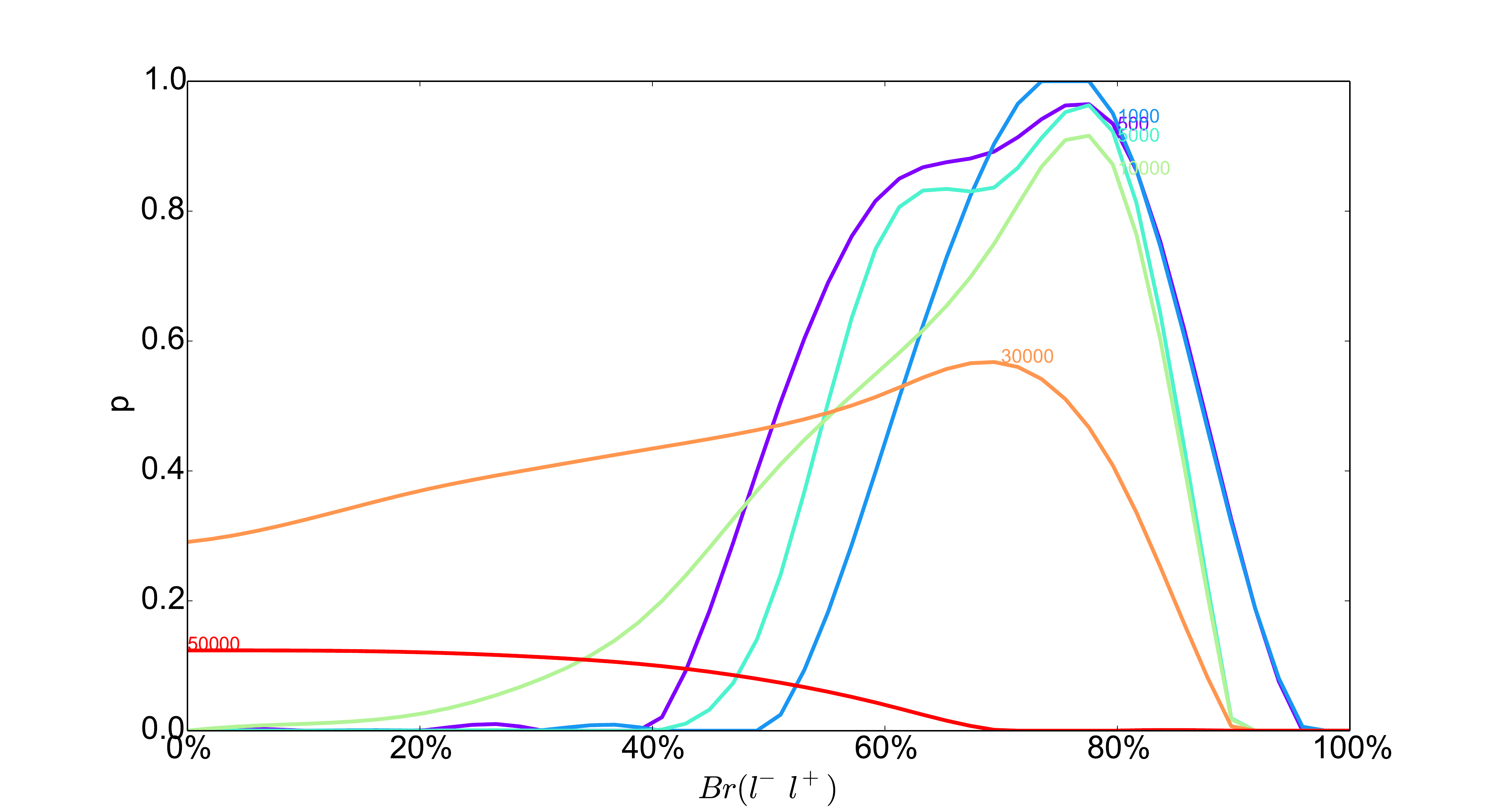}
\caption{ \small \label{nfw_med}  
Fitting results of the multi-annihilation channel DM
scenario to AMS-02 $\frac{e^+}{e^-+e^+}$ data
for the combination of {\it NFW-MED}. Description goes from top to bottom and
from left to right.
i) Upper-left: the allowed region in the plane of 
 $(M_{\chi}$, $\left\langle \sigma v 
 \right\rangle )_{tot}$. The blue region indicates 
 $0.32 < p{\rm -value} < 1.0$ (68.3\% CL), the green
region $0.05 < p{\rm -value} < 0.317$ (95\% CL), and the yellow
region $0.01 < p-{\rm value} < 0.05$ (99\% CL).
ii) Upper-right: ($M_{\chi}$, BR). The branching ratio of each
channel for the best-fit points, labelled on the left $y$-axis. 
The blue-shaded region is the
$p-$value of the best-fit points labelled on the right $y$-axis.
iii) Middle-left: ($M_{\chi}$, Br). Same as ii), but for
summing over the leptonic, quark, and massive gauge boson channels. 
%
iv) Middle-right: ($M_{\chi}$, $\chi^2$). Each of the 
solid curves shows the $\chi^2$ in the fitting to the AMS-02 
$\frac{e^+}{e^-+e^+}$ data using only a single DM annihilation channel, 
namely, $b\bar b$, $t\bar t$, $e^- e^+$, $\mu^- \mu^+$, $\tau^- \tau^+$,
$ZZ$, or $W^- W^+$.
The dashed curve represents the $\chi^2$ from the multi-annihilation
channel DM scenario.
v) Lower: (${\rm Br}(l^-l^+),\, p$). The sum of leptonic branching ratios
${\rm Br}(l^-l^+)\equiv\sum_{f=e,\mu,\tau}{\rm Br}(f\bar{f})$ vs the $p$-values
for a number of $M_{\chi}$ values,
labelled right next to the curves  while marginalizing all other
fitting parameters.  
}
\end{figure}

\begin{figure}[th!]
\centering
\includegraphics[width=3.2in]{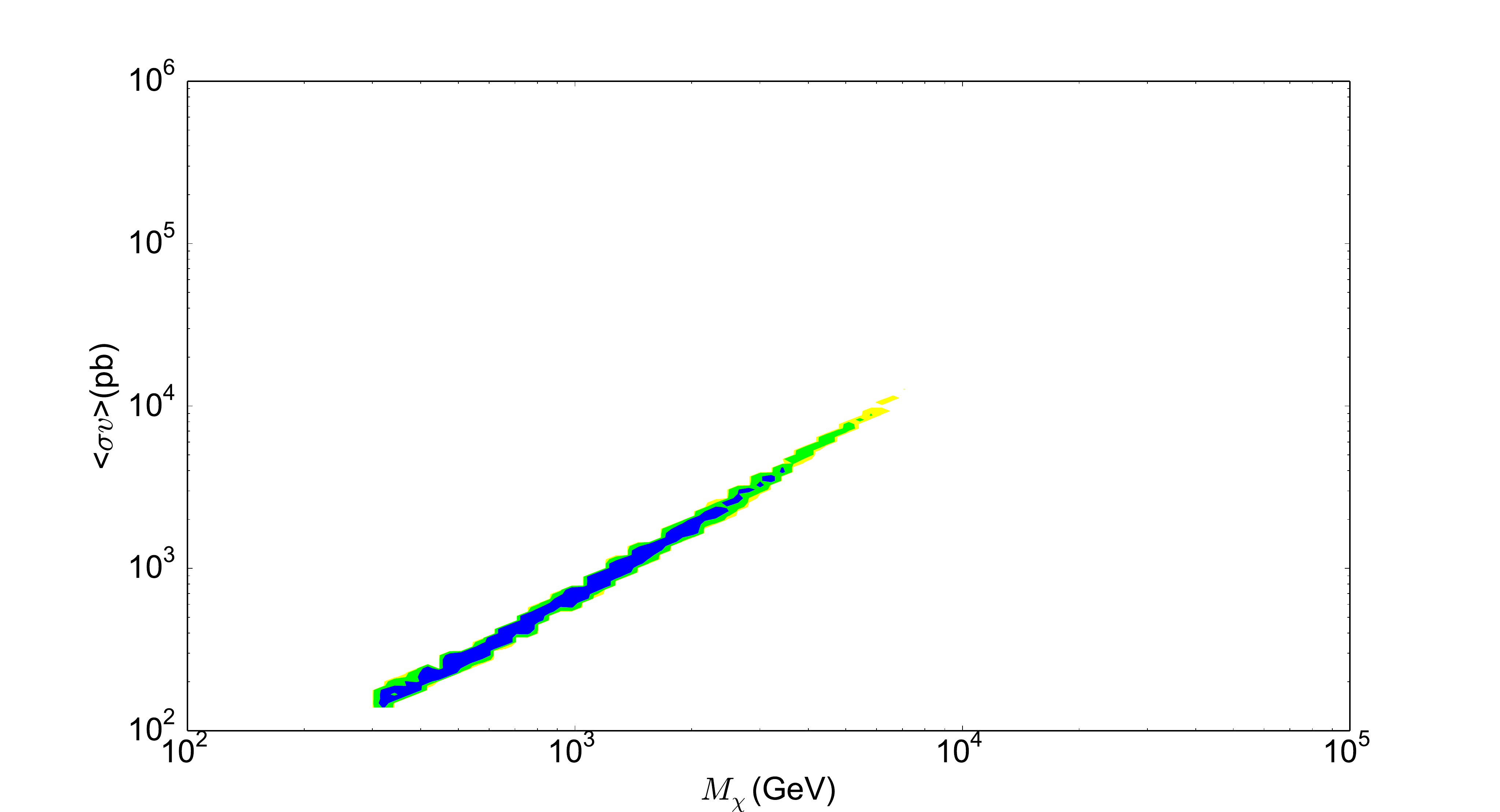}
\includegraphics[width=3.2in]{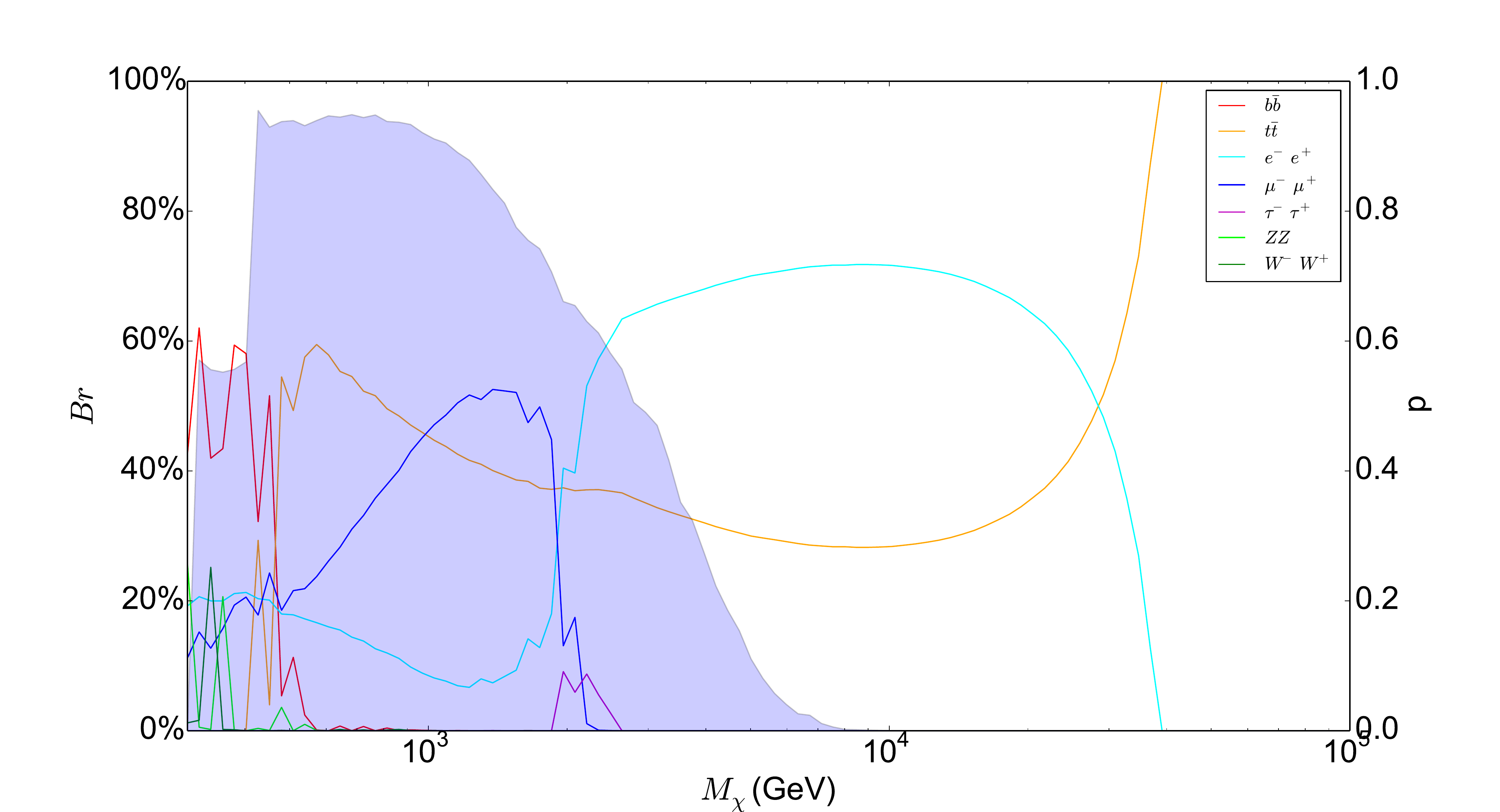}
\includegraphics[width=3.2in]{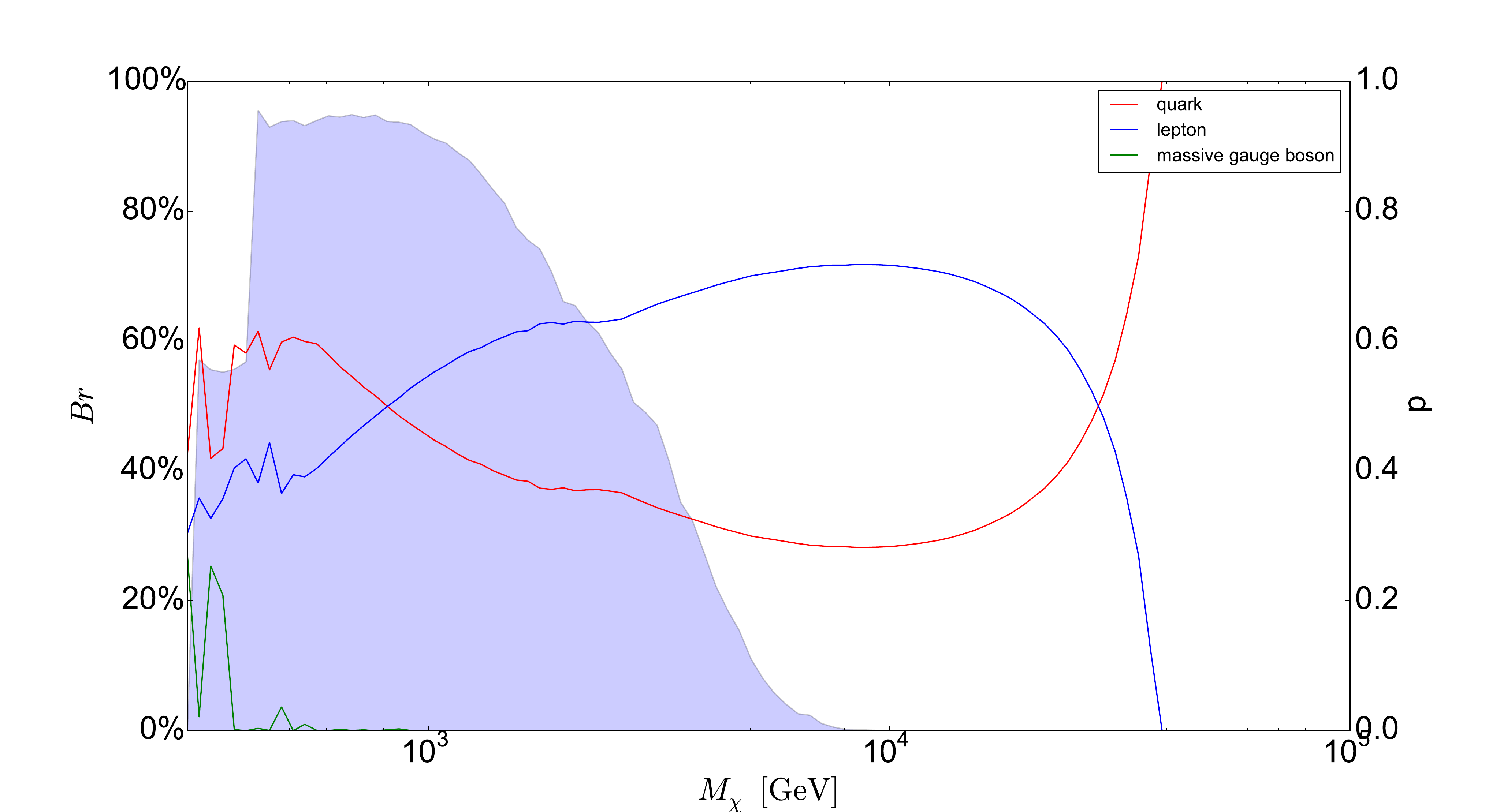}
\includegraphics[width=3.2in]{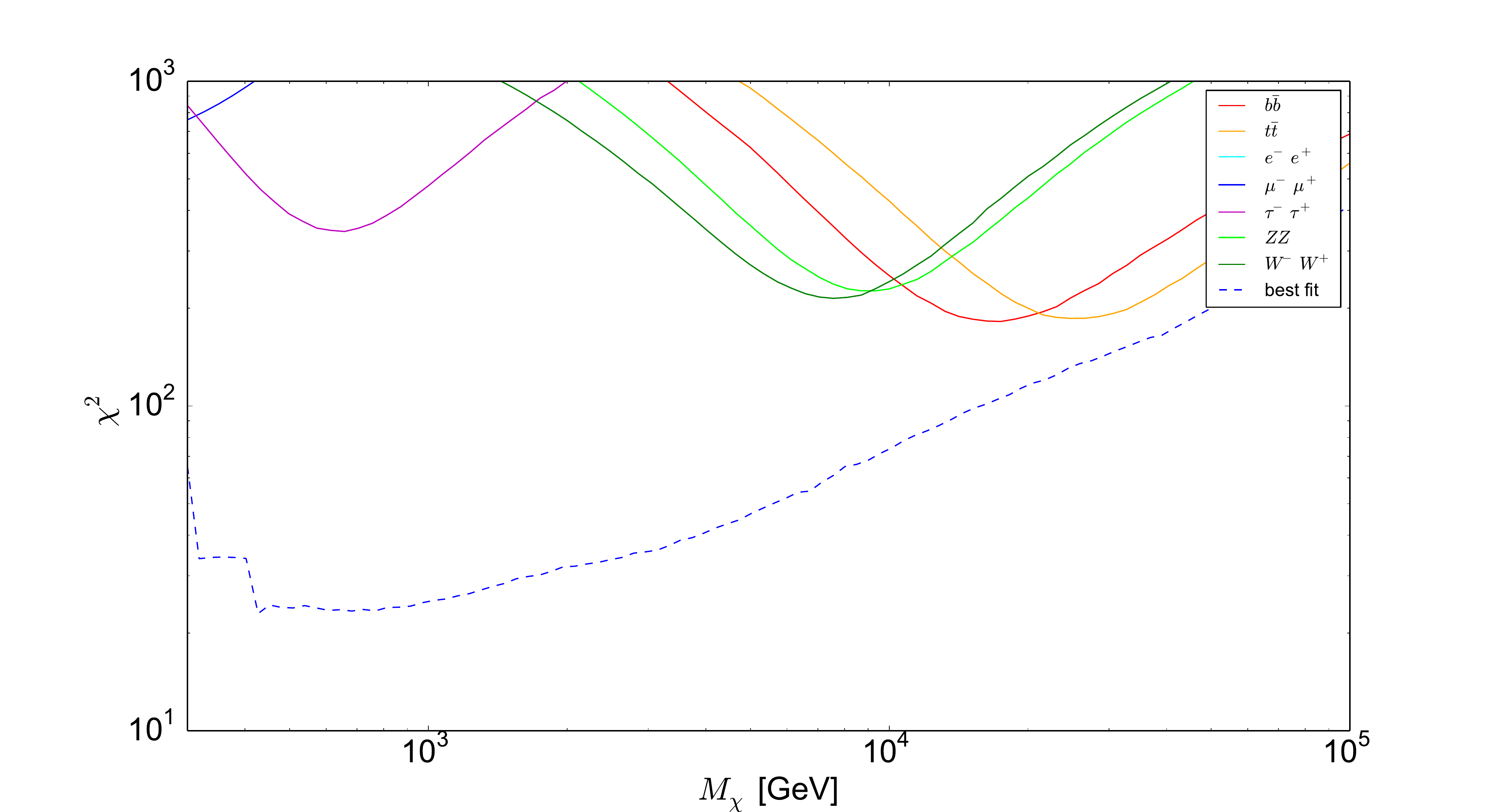}
\caption{ \small \label{nfw_max} 
The same as Fig.~\ref{nfw_med} but for the combination of of NFW-MAX, except
without the last panel.
}
\end{figure}

\begin{figure}[th!]
\centering
\includegraphics[width=3.2in]{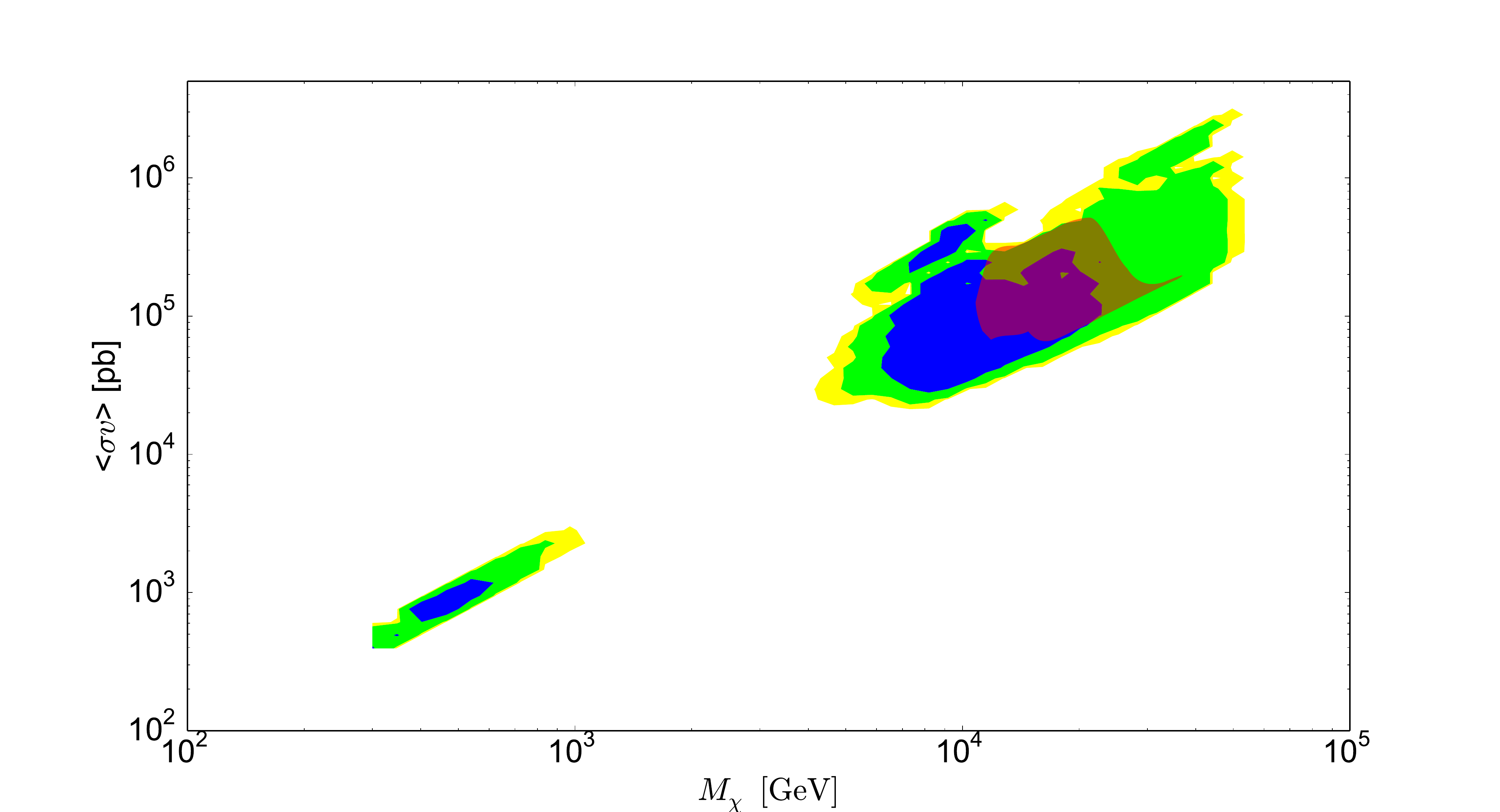}
\includegraphics[width=3.2in]{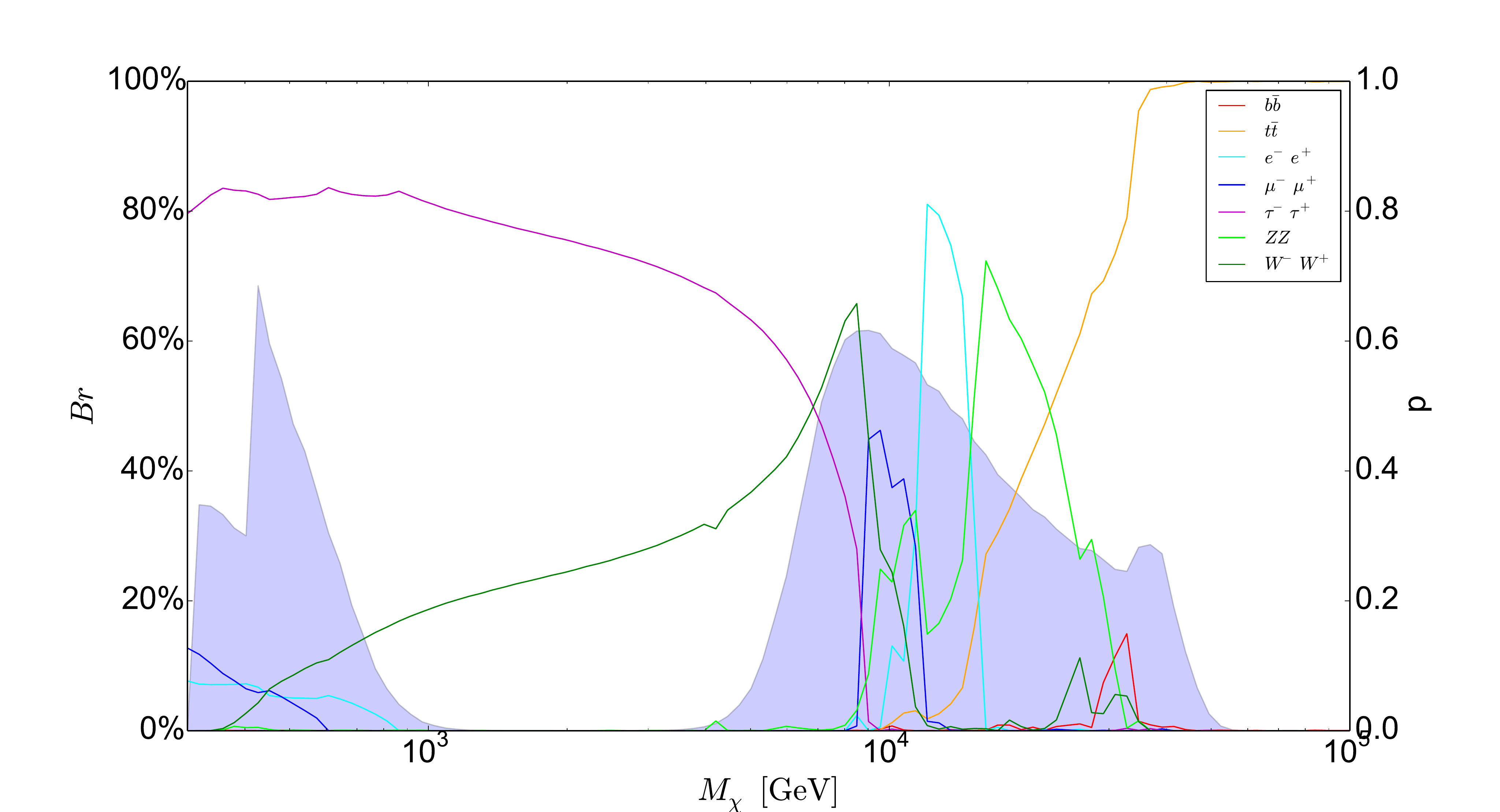}
\includegraphics[width=3.2in]{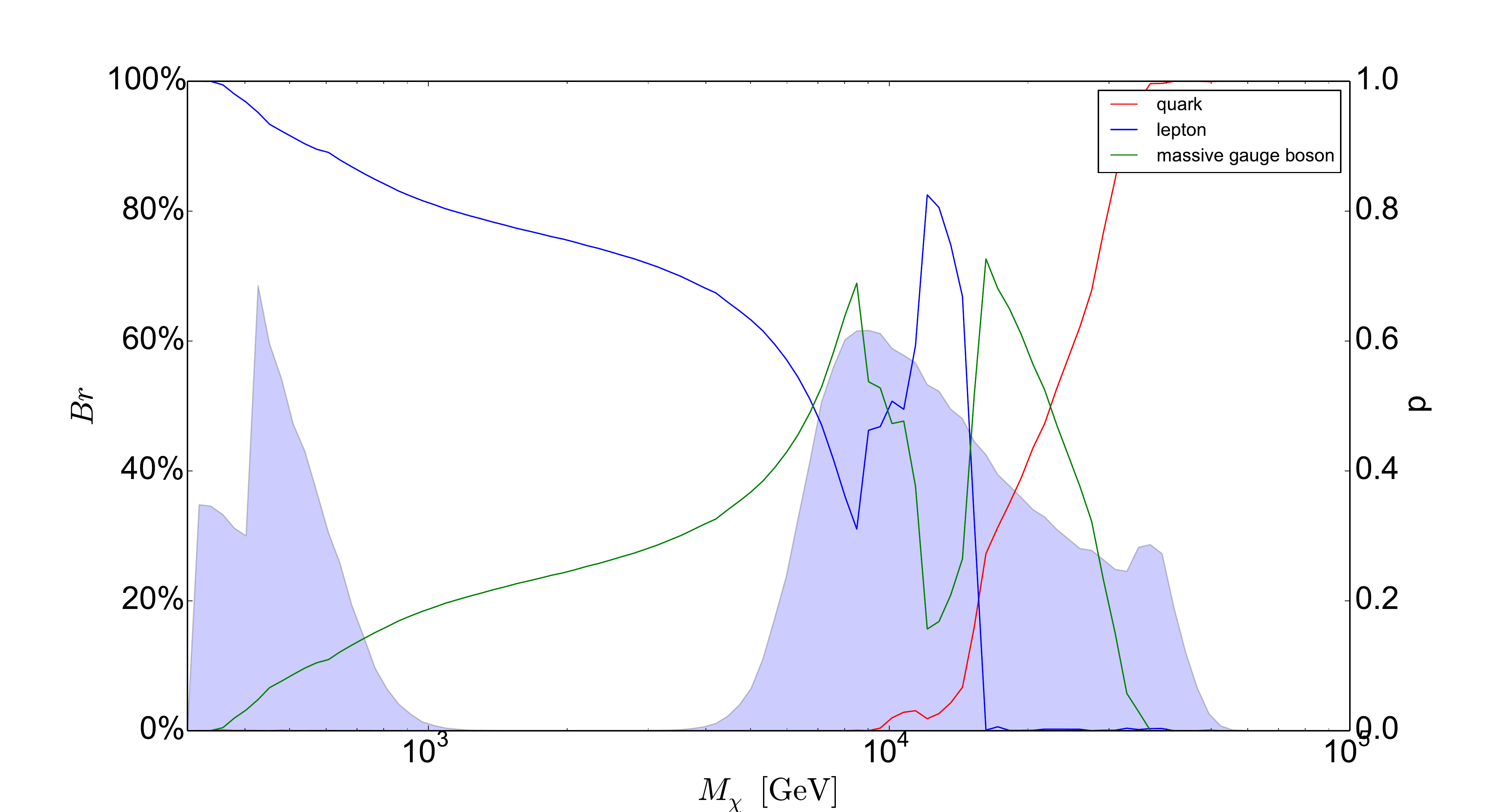}
\includegraphics[width=3.2in]{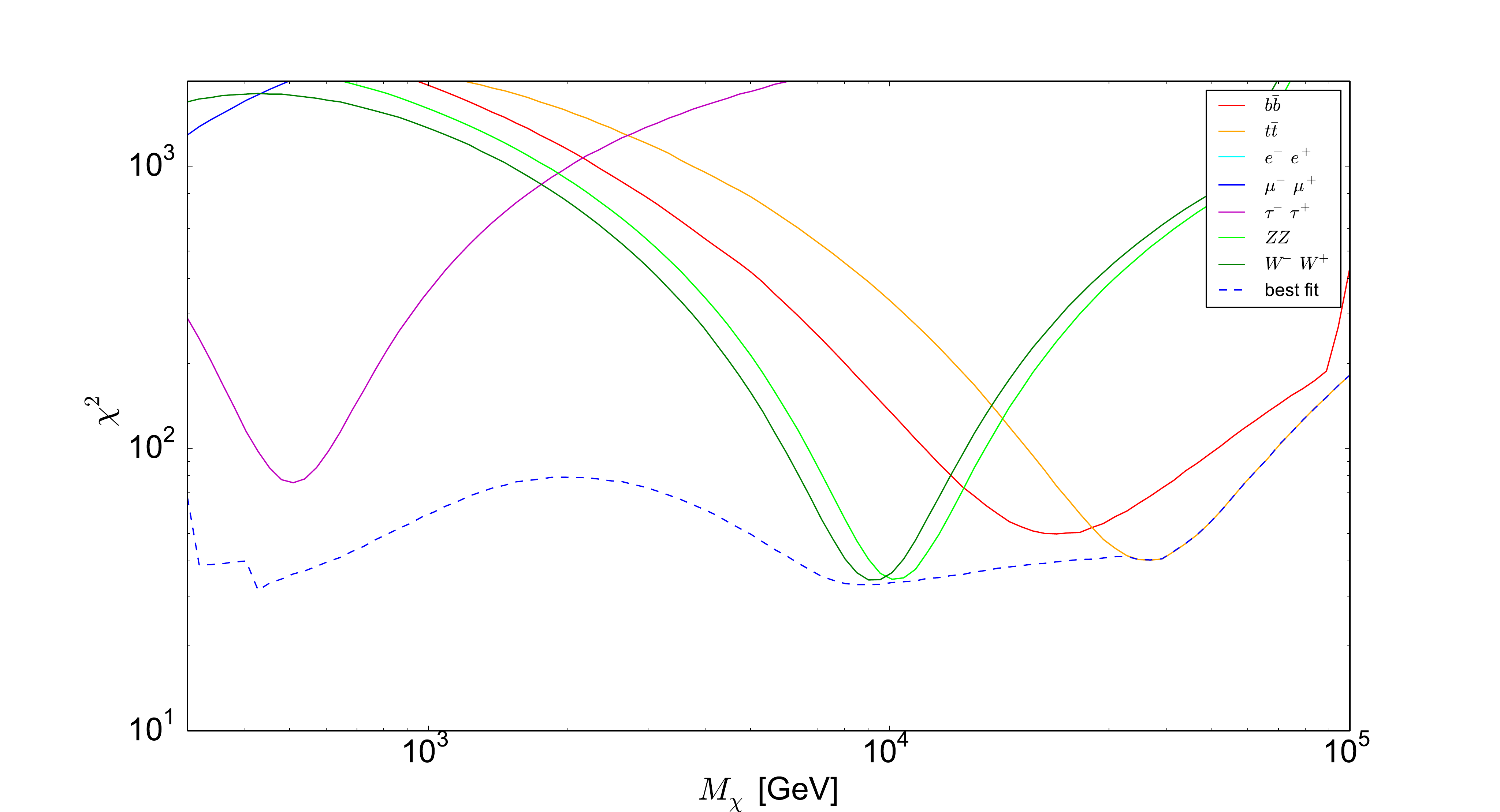}
\caption{ \small \label{nfw_min} 
The same as Fig.~\ref{nfw_med} but for the combination of of NFW-MIN, except
without the last panel.
The compensating color in upper-left panel in the plane of 
($M_{\chi}$, $\left\langle \sigma v \right\rangle$) 
shows the allowed region of fitting simultaneously to both
AMS-02 $\frac{e^+}{e^-+e^+}$ and $\bar{p}/p$ data 
with $p$-value $> 0.05$ for each dataset.
}
\end{figure}

\begin{figure}[th!]
\centering
\includegraphics[width=3.2in]{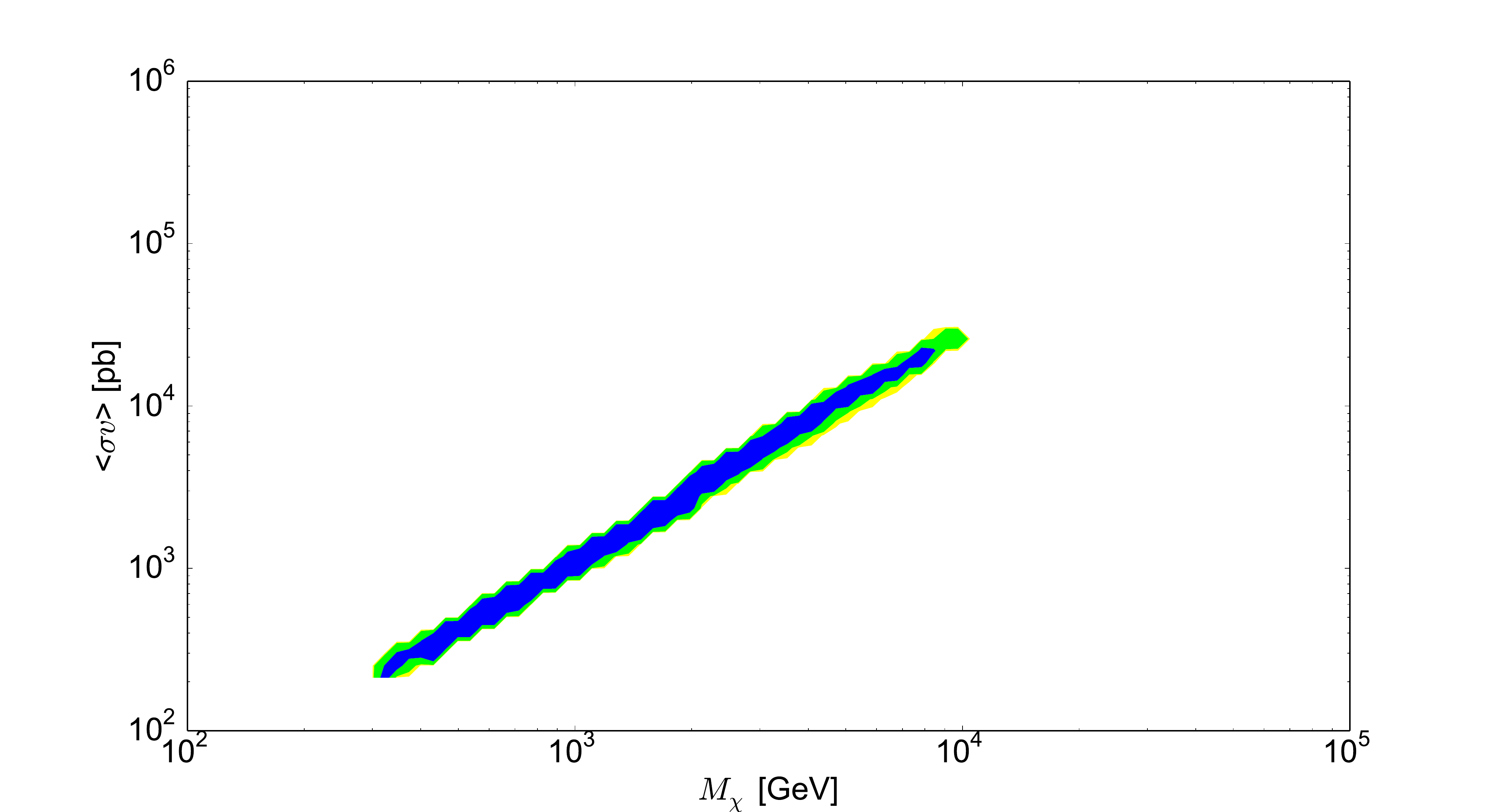}
\includegraphics[width=3.2in]{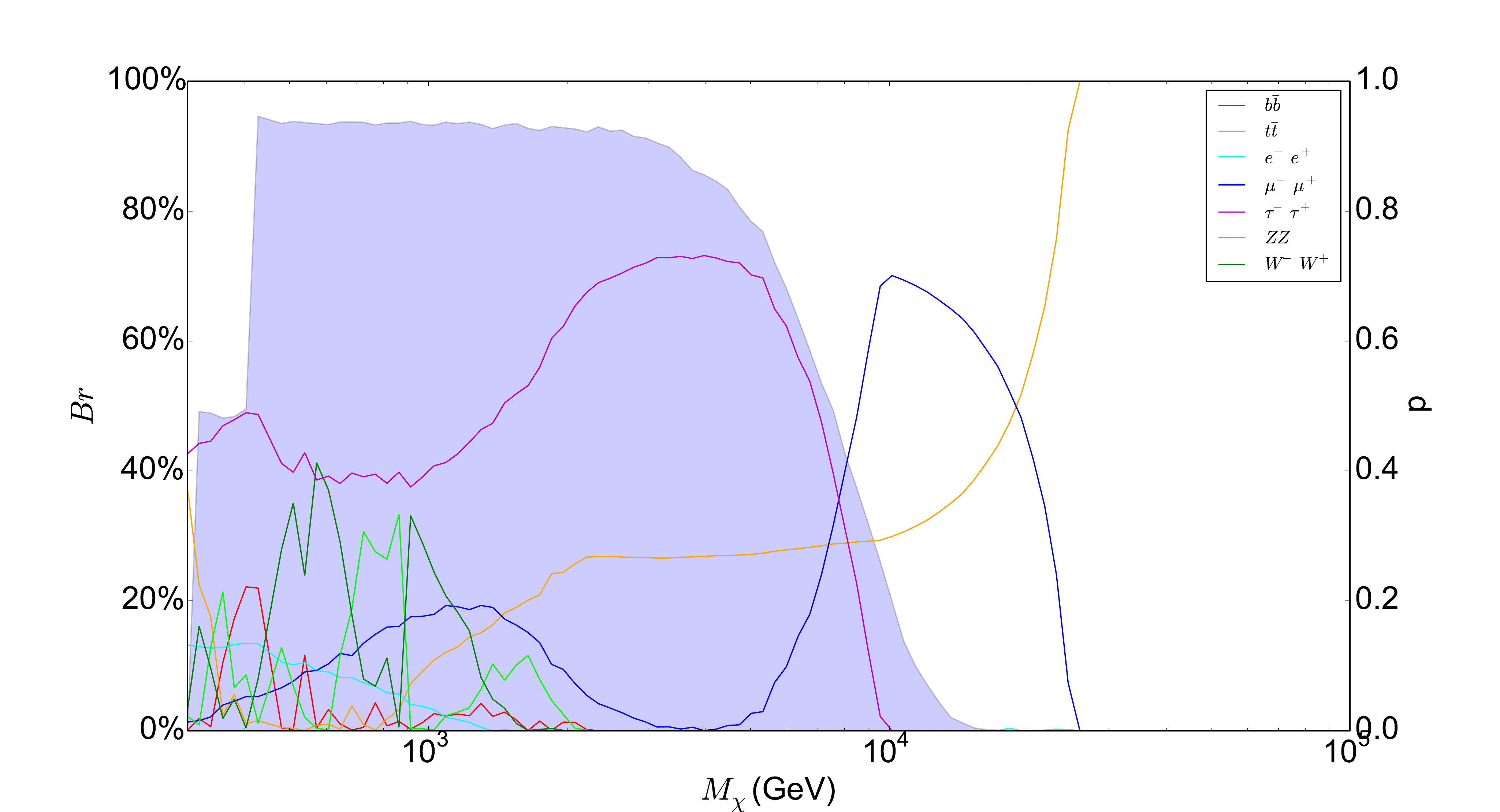}
\includegraphics[width=3.2in]{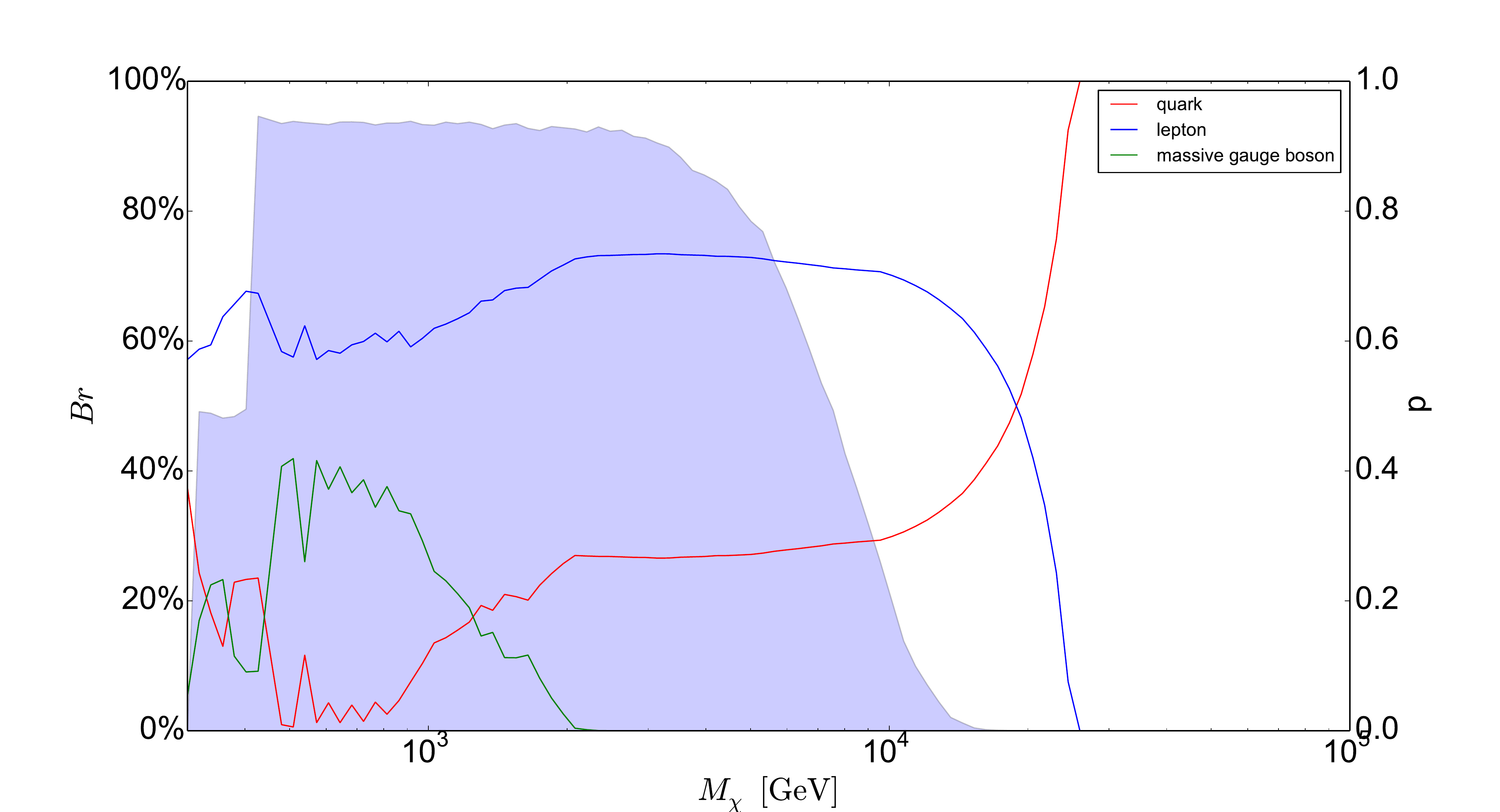}
\includegraphics[width=3.2in]{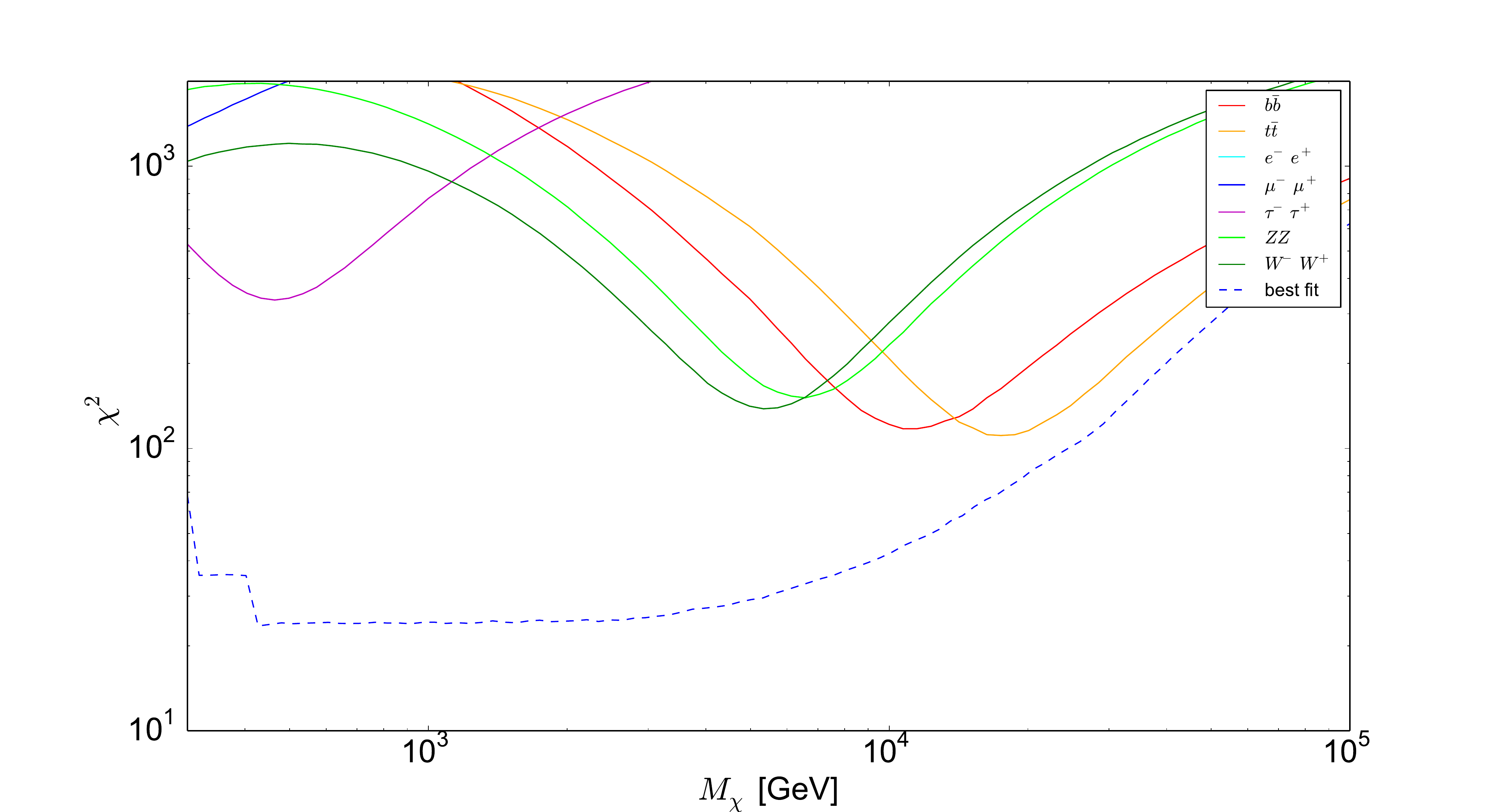}
\caption{ \small \label{nfw_dc} 
The same as Fig.~\ref{nfw_med} but for the combination of of NFW-DC, except
without the last panel.
}
\end{figure}

We use the $\chi^2$ and $p$-value as our statistical measures to 
describe the deviation of our model from the data and the goodness of the fit,
respectively.
\footnote
{Assuming the goodness-of-fit statistics follows $\chi^2$ statistics,
the $p$-value for the hypothesis is given by \cite{pdg}
\[
 p = \int_{\chi^2}^{\infty} f(z;n) dz
\]
where $n$ is the degrees of freedom and 
\[
 f(z;n) = \frac{z^{n/2-1} e^{-z/2} }{ 2^{n/2} \Gamma(n/2) }\,.
\] 
} 
Our scanning strategy is fixing certain values of $M_{\chi}$ and 
$\left\langle \sigma v \right\rangle_{tot}$,  meanwhile allowing the
branching ratio of each channel to float as long as they satisfy the
sum rule to obtain the minimal $\chi^2$ or maximal
$p$-value. Repeating the whole procedures to scan the parameter space of
$M_{\chi}$ and $\left\langle \sigma v \right\rangle_{tot}$. The results
are shown in Figs.~\ref{nfw_med},\ref{nfw_max},\ref{nfw_min}, and 
\ref{nfw_dc}, corresponding to MED, MAX, MIN, and DC diffusion models, 
respectively.

In the panel of ($M_{\chi}$, $\left\langle \sigma v \right\rangle$) in
each of the 
Figs.~\ref{nfw_med},\ref{nfw_max},\ref{nfw_min}, and \ref{nfw_dc}, we
show the 68.3\% (blue, $0.32 < p-{\rm value} < 1.0$) 95\% (green,
$0.05 < p-{\rm value} < 0.317$), and 99\% (yellow, $0.01 < p-{\rm
  value} < 0.05$) as regions of confidence level (CL) of the
best-fitting points with varying branching ratios of the DM
annihilation channels.
In these panels, independent of the diffusion models, 
they show the narrow window in 
$\left\langle \sigma v \right\rangle_{tot}$ restricted by AMS-02
$\frac{e^+}{e^++e^-}$ data for each value of $M_{\chi}$, 
despite of the variation of seven annihilation channels.
Also, the total annihilation cross section $\left\langle \sigma v
\right\rangle_{tot}$ strongly correlates with and is
roughly proportional to the $M_{\chi}$, but it is much
larger than the thermal DM cross section, which is about $\left\langle
\sigma v \right\rangle_{thermal}\simeq1\;\rm{pb}\;\simeq\;3\times
10^{-26} cm^3/s$.

In the panel of ($M_{\chi}$, Br) in each of the
Figs.~\ref{nfw_med},\ref{nfw_max},\ref{nfw_min}, and \ref{nfw_dc}, the
light blue-shaded area shows the $p$-values of the best-fit point for
the corresponding $M_{\chi}$, and the $p-$value is labelled on the
right-hand y-axis of the panel.
It manifests itself in the ($M_{\chi}$, Br) panel that
once the DM mass is below 300 GeV, the $p$-value is always less than $0.05$,
independent of the diffusion models, and no matter how the combination 
of the annihilation channels changes. 
In other words, we can rule out the DM mass below 
300 GeV as an explanation for the AMS-02 $\frac{e^+}{e^++e^-}$
excess at $95\%$ CL. 
This is because the last two AMS-02 $\frac{e^+}{e^++e^-}$ data points
with the highest energy between 260 GeV to 500 GeV 
cannot be explained by DM annihilation for the DM mass below
300 GeV, due to the cliff shape of the positron energy spectrum from DM
annihilation around $M_{\chi}$. 
On the other hand, the upper limit for the DM mass, 
which are diffusion model dependent, are 5.5, 60, 50, and 13 TeV for 
MAX, MED, MIN, and DC, respectively.

There are two panels of ($M_{\chi}$, Br) in each of the 
Figs.~\ref{nfw_med},\ref{nfw_max},\ref{nfw_min}, and \ref{nfw_dc}, one of
them shows the branching ratio of individual channel for the best-fit
point at the corresponding $M_{\chi}$, while the other one shows the 
branching-ratio sum over the leptons($e^+e^-+\mu^+\mu^-+\tau^+\tau^-$),
quarks($b\bar{b}+t\bar{t}$), or massive gauge-boson($W^+W^-+ZZ$)
channels.
When the DM mass is below 10 TeV, the leptonic channels dominate the
branching ratio ($\textrm{BR}({\rm leptons})\gtrsim 60\%$), except for
the region of $M_{\chi}<$ 1 TeV in the MAX diffusion model, where the
quark channels are able to compete with leptonic
channels. Nonetheless, among these four diffusion models, the quark
and gauge-boson channels still play non-negligible roles because 
($\textrm{BR}({\rm gauge\,bosons})+\textrm{BR}({\rm quarks}) \gtrsim
20\%$).
Since $\left\langle \sigma v \right\rangle_{tot}\sim \cal{O}$ $(10^3 \sim
10^4 \rm{pb})$, even $10\%$ of the $W^+W^-,\;ZZ$, and $b\bar{b}$ channels
would generate enormous antiproton flux, and that is why it is severely
constrained by the antiproton data~\cite{pamela_antip}. 
Therefore, the largest $p$-value point for each
$M_{\chi}$ would be ruled out by the antiproton data.
Among these four diffusion models, only the MIN diffusion model at the
region with $M_{\chi}<$ 350 GeV, where the three leptonical channels
can fit the AMS-02 $\frac{e^+}{e^++e^-}$ with $p-$value larger than
0.05.
Another feature is that once the DM mass goes above 30 TeV, the
$t\bar{t}$ channel becomes more and more important and eventually
becomes the major channel, because of the energy spectrum of
$t\bar{t}$ is relatively softer than the other channels.

In the following, we consider how the AMS-02
$\frac{e^+}{e^++e^-}$ data can distinguish among different
channels. The panel of ($M_{\chi}$, $\chi^2$) in each of the 
Figs.~\ref{nfw_med},\ref{nfw_max},\ref{nfw_min}, and \ref{nfw_dc} shows
the chi-squares minimum with respect to $M_{\chi}$ if the DM can only
annihilate into a single channel.  The solid lines represent the
single-annihilation channel scenario, while the dashed line shows the
chi-square minimum of multi-annihilation channel scenario.
We could see that the DM with multi-annihilation channel scenario
can reduce the $\chi^2$ significantly, although more parameters
are involved. When the DM mass is less than 1 TeV, 
the $\tau^+\tau^-$ single-annihilation channel scenario can provide 
a relatively smaller value of $\chi^2$ than the other
channels, and the minimum $\chi^2$ occurs at around 
$M_{\chi}=$650, 800, 500, and 450 GeV for MAX, MED, MIN, and DC 
diffusion models, respectively.
The two massive vector-boson channels attain their $\chi^2$ minimum at about
$M_{\chi}=7$, 10, 10, and 5 TeV for MAX, MED, MIN, and DC diffusion models, 
respectively, because they generate moderate positron energy
spectra in between the leptonic and quark channels. 
The $t\bar{t}$ channel, on the other hand, generates a further softer 
positron energy spectrum, such that
the $\chi^2$ minimum occurs at around $M_{\chi}=25$, 40, 35, and 20 TeV, 
for the MAX, MED, MIN, and DC diffusion models, respectively.
Once $M_{\chi} > 20$ TeV, the curve of $t\bar{t}$ is in unison with
the blue-dash curve, and the $t\bar{t}$ channel takes the full
responsibility for fitting the data, and other channels become
irrelevant.
{}From these results, we can see that the lepton channels are 
well separated from quark and massive gauge boson channels, and 
therefore we anticipate that the
AMS-02 $\frac{e^+}{e^++e^-}$ data is able to distinguish the leptonic
channels from the other channels in the DM multi-annihilation channel
scenario.  However, the $\chi^2$ minima of $W^+W^-$ and $ZZ$ are
almost overlapping, hence it is hard to distinguish them from each
other.
In the bottom-left panel of Fig.~\ref{nfw_min}, 
with 300 GeV $\lesssim M_{\chi}\lesssim$ 450 GeV, 
pure leptonic channels are solutions of AMS-02 positron fraction data 
under MIN diffusion model and NFW profile.

For MIN diffusion model in Fig.~\ref{nfw_min}, there is no
contradiction between the panels of ($M_{\chi}$, Br) and ($M_{\chi}$,
$\chi^2$) around $M_{\chi}=$ 12 TeV. At the first glance, the panel of
($M_{\chi}$, $\chi^2$) points out that the $WW$ and $ZZ$ channels are the
most important ones to fit the data at this $M_{\chi}$. However, the
${\rm Br}(e^+e^-)$ is the largest from the panel of ($M_{\chi}$, Br). The
reason is that the $e^+e^-$ channel almost decouples at this DM mass
region, and so the positron spectrum from it is way above the energy scale of
AMS-02 $\frac{e^+}{e^++e^-}$ data, such that dramatically change of
$\left\langle \sigma v \right\rangle_{\chi\bar{\chi}\rightarrow
  e^+e^-}$ does not influence the fitting too much. Therefore, the $WW$ and
$ZZ$ channels are still the major channels to fit the data 
around $M_{\chi}=$ 12 TeV.

The panel of (${\rm Br}(l^+l^-)$, p) in Fig.~\ref{nfw_med} shows 
how many percentages of the leptonic 
channels are required to fit to the AMS-02 $\frac{e^+}{e^++e^-}$
data with the corresponding $M_{\chi}$
in the DM multi-annihilation scenario, while marginalizing all the
massive gauge boson ($VV$) and quark ($q\bar{q}$) channels.
%
Take $M_{\chi}$= 1000 GeV, for example, the $\chi^2$
minimum occurs with ${\rm Br}(l^+l^-)\approx 75^{+20}_{-23}\%$ and 
${\rm Br}(q\bar{q})+{\rm Br}(VV)\approx 25^{+23}_{-20}\%$ at 95\% CL
in the multi-annihilation scenario, while 
the single-annihilation channel scenario is the $\tau^+\tau^-$ 
channel but at a much larger $\chi^2$.
Two inferences we can
draw from here for $M_{\chi}$= 1000 GeV: 
(i) the major channels are the leptonic, and (ii)
the three lepton channels alone, ${\rm Br}(l^+l^-)= 100\%$, could not fit
well to the data. Instead, they need help from softer positron
spectra of quarks or gauge bosons.
When $M_{\chi}<$ 10 TeV, the three leptonic channels dominate. 
On the other hand, for $M_{\chi}>$ 10 TeV, the $VV$ and $q\bar{q}$ 
channels gradually become the major contributions.
For instance, when $M_{\chi}=$ 30 TeV, ${\rm 
  Br}(l^+l^-)\approx 70^{+18}_{-70}\%$.  Yet, without any lepton channels
the $VV$ and $q\bar{q}$ still have $p$-value = 0.3. In another words,
the AMS-02 $\frac{e^+}{e^++e^-}$ data constrain more severely on the
branching ratio of leptonic channels for $M_{\chi}<$ 10 TeV.

\begin{figure}[th!]
\centering
\includegraphics[width=5.4in]{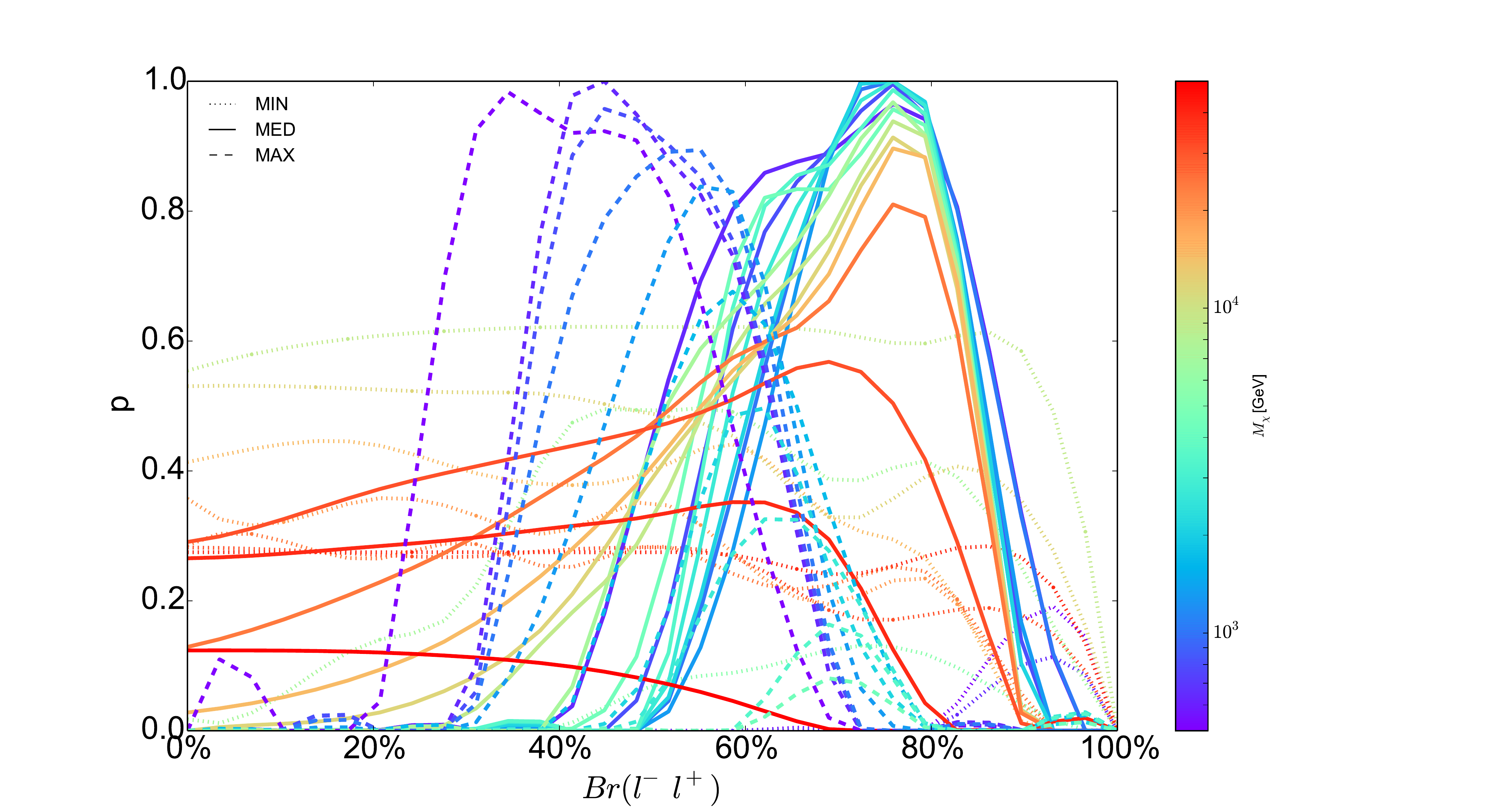}
\caption{ \small \label{br_all} 
The same description as the lower panel in Fig.~\ref{nfw_med}, 
except for using all different diffusion models.  
MIN: dotted curves; 
MED: solid curves; 
MAX: dashed curves. 
The color bar on the right-hand side indicates the DM mass $M_{\chi}$.
}
\end{figure}

In Fig.~\ref{br_all}, we would like to demonstrate how different background
of $\frac{e^+}{e^++e^-}$ originating from different diffusion models
will modify the conclusions that we made in previous paragraph. 
For all
three diffusion models MAX, MED, and MIN at $M_{\chi}\simeq$ 1 TeV,
the values of ${\rm Br}(l^+l^-)$ are still confined in a narrower
window with about $\pm 20\%$ deviation at 95\% CL from the
best-fit value of ${\rm Br}(l^+l^-)$. 
On the other hand, for $M_{\chi}>$ 10 TeV
the confinement is much loose such that the values of
${\rm Br}(l^+l^-)$ can go down to zero in MED and MIN cases. 
Another feature we can see from this figure is that in the MIN (MAX) diffusion
model, the values of ${\rm Br}(l^+l^-)$ concentrate at slightly
larger (smaller) values than those of the MED model.

\section{AMS-02 antiproton-proton ratio data and fitting results}

During 15-17 April 2015 at CERN Geneva, the AMS collaboration
presented the preliminary results on the spectrum of the
antiproton-to-proton ratio for proton kinetic energy (KE) up to 450
GeV \cite{ams02_antip}, with much higher precision than other
measurements in the past.  The data points with KE between 20 and 450
GeV show a flat structure.  Comparing the AMS-02 and PAMELA
$\bar{p}/p$ data, with KE larger than 5 GeV up to 180 GeV, these two
datasets are consistent within the errors.

\begin{figure}[th!]
\centering
\includegraphics[width=5.4in]{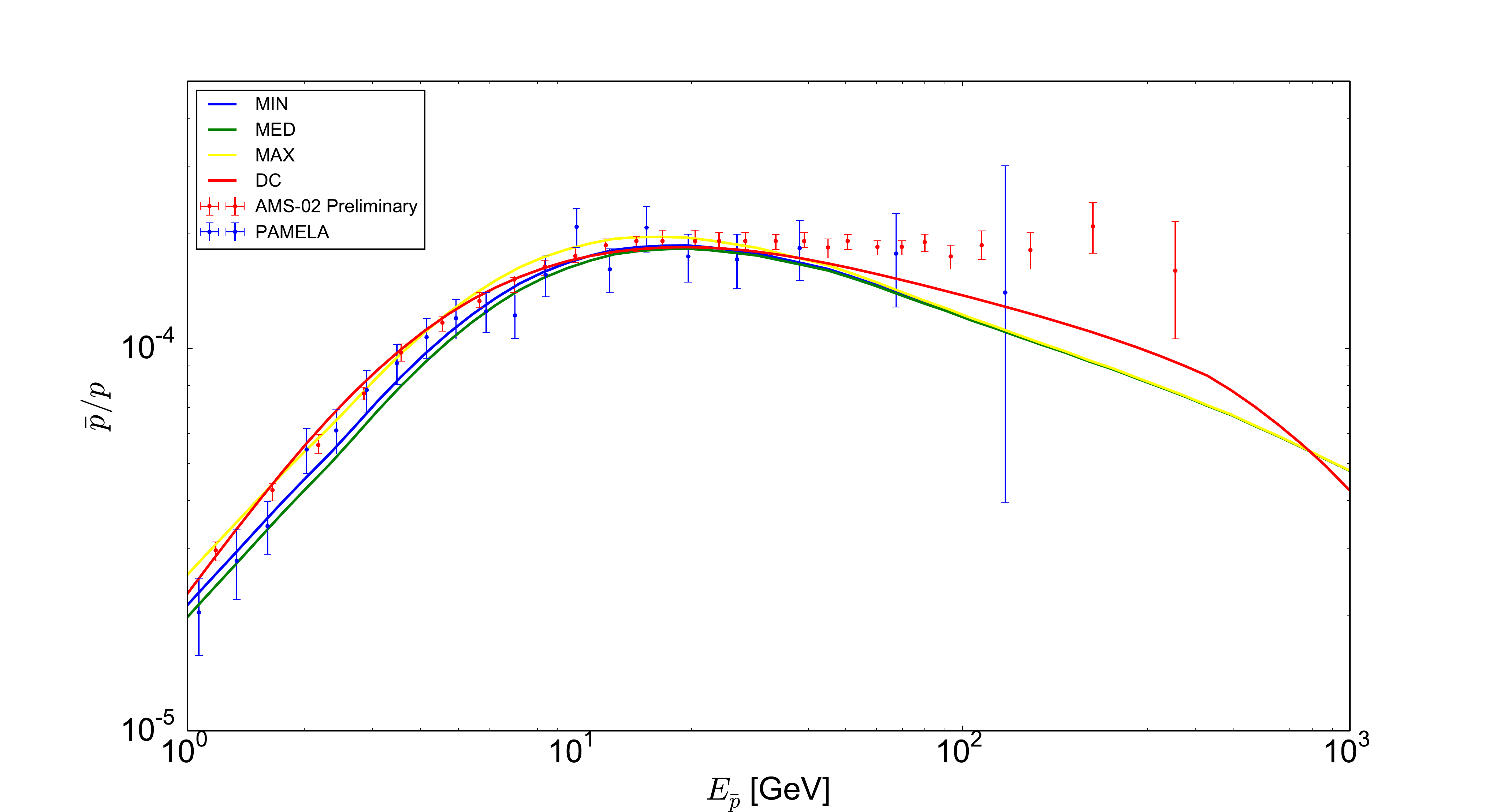}
\caption{ \small \label{antip_bkgd} 
Astrophysical backgrounds for the observable $\bar{p}/p$ data with the
MIN, MED, MAX, and DC diffusion models. 
The red dots are the AMS-02 $\bar{p}/p$ preliminary data. 
The blue dots are the PAMELA $\bar{p}/p$ data.
}
\end{figure}

The derivation of the antiproton background is analogous to the positron
case.  First, we specify the diffusion model and determine the proton
flux from the data, then compute the antiproton flux based on
Galprop~\cite{galprop1,galprop2}. Various backgrounds originating from
the corresponding diffusion models are shown in
Fig.~\ref{antip_bkgd}. There is a moderate excess of the AMS-02
$\bar{p}/p$ data above various backgrounds. Comparing among these four
backgrounds, the one from the DC diffusion model provides a slight better
fit to the AMS-02 $\bar{p}/p$ data.

\begin{figure}[th!]
\centering
\includegraphics[width=3.2in]{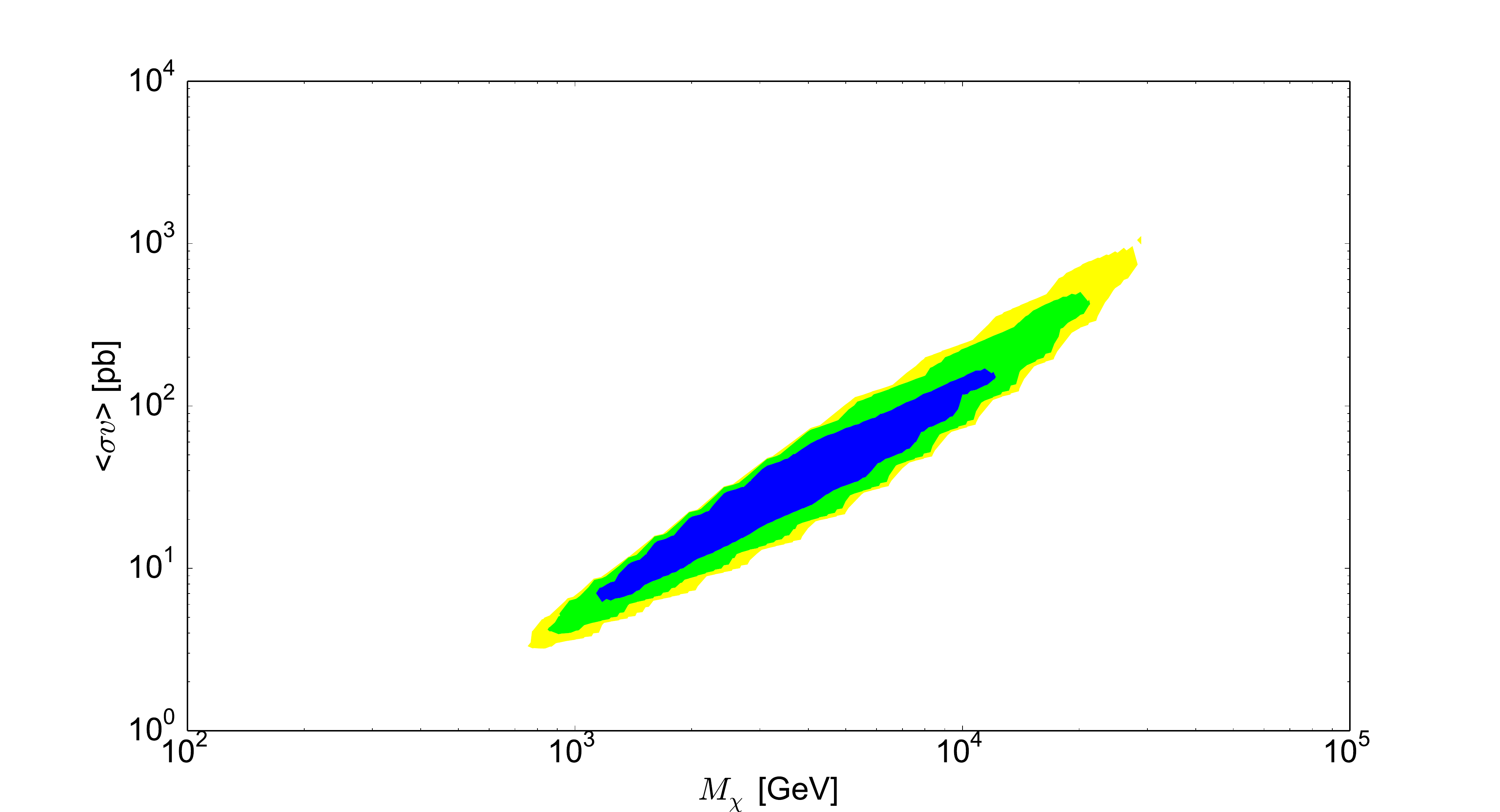}
\includegraphics[width=3.2in]{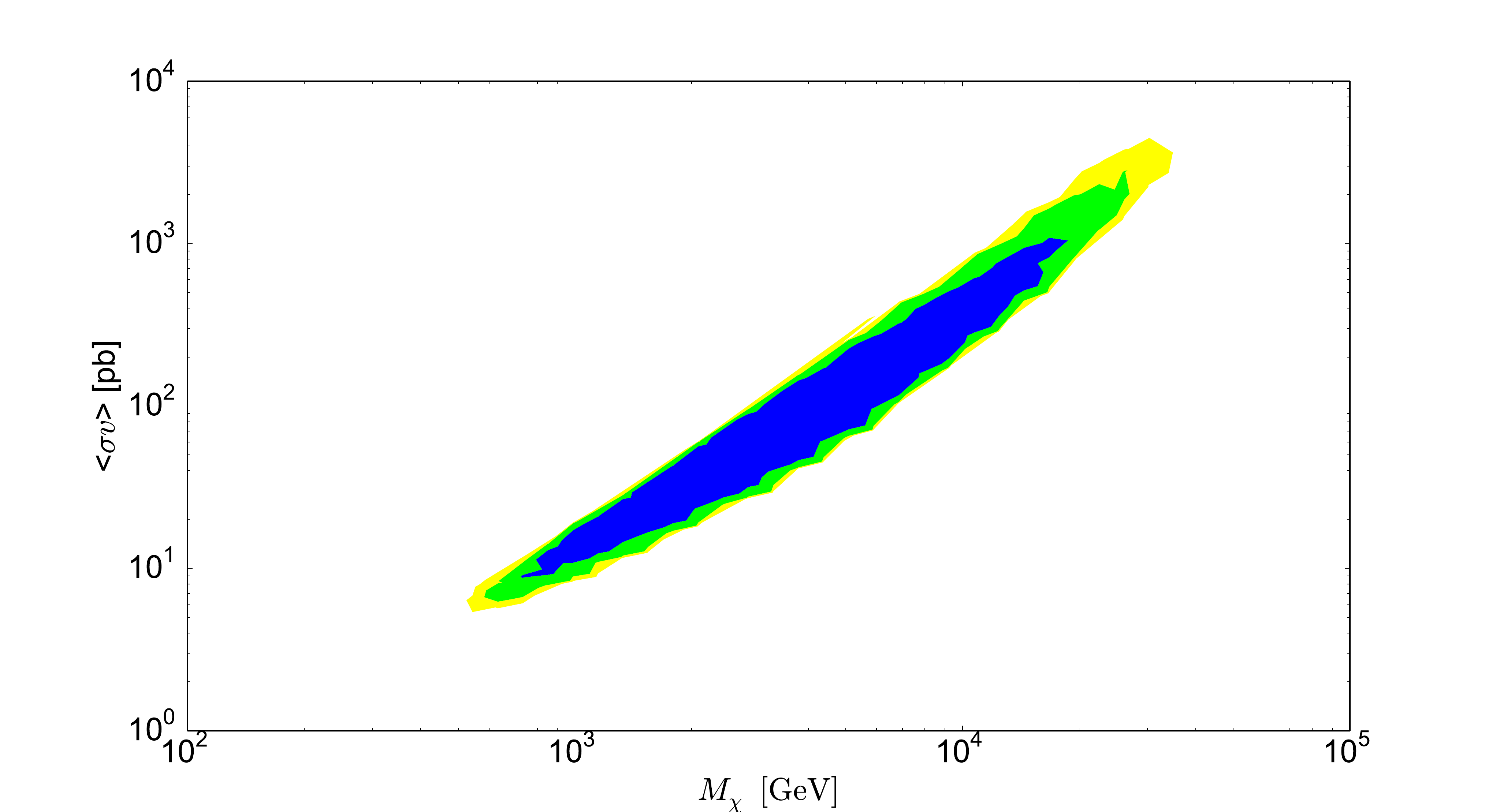}
\includegraphics[width=3.2in]{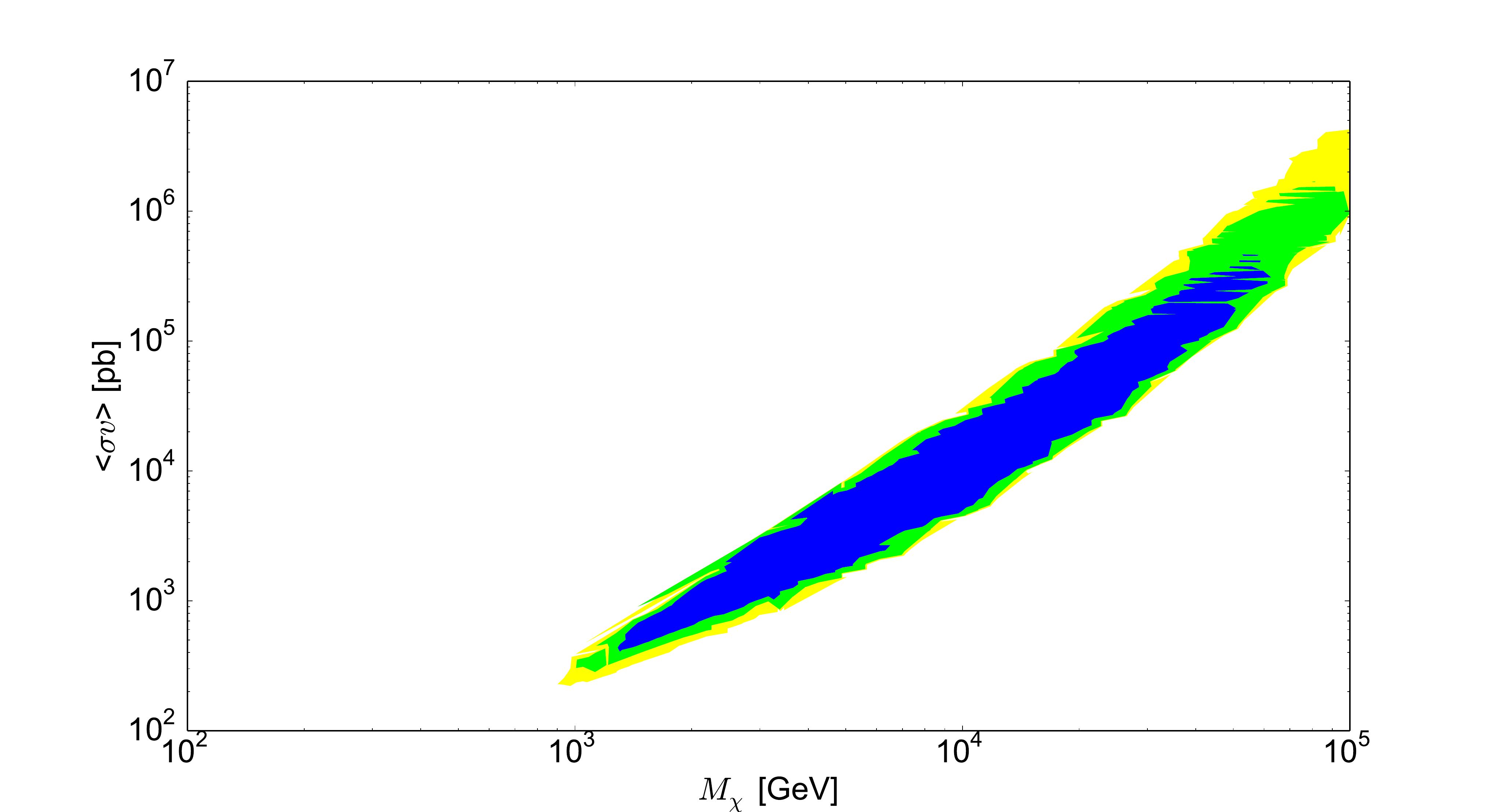}
\includegraphics[width=3.2in]{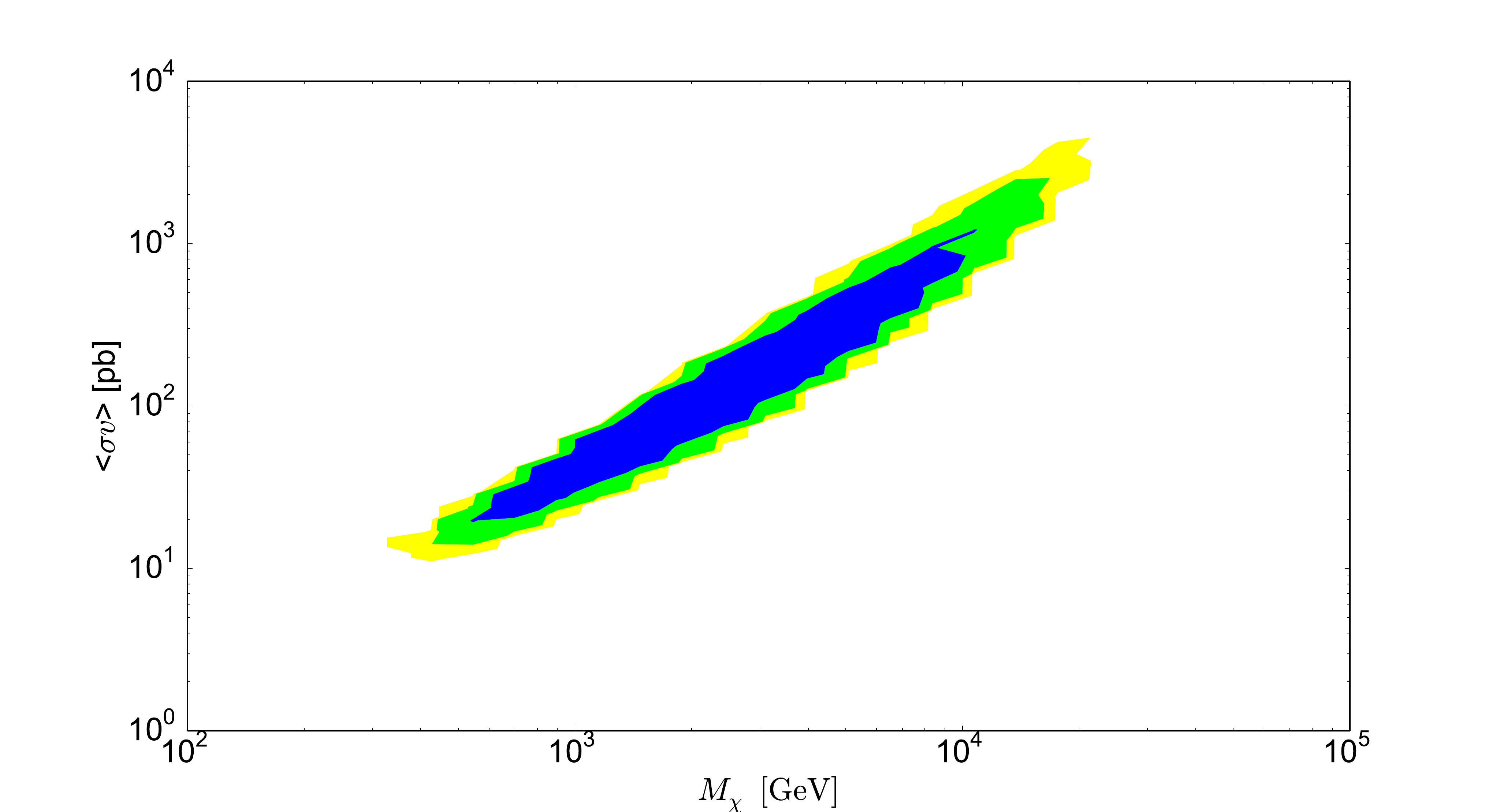}
\caption{ \small \label{antip_fit} 
The fitting results in the plane of $(M_{\chi},\, \langle \sigma v \rangle )$
of the multi-annihilation channel DM scenario 
to the AMS-02 $\bar{p}/p$ data. 
The relevant channels are $t\bar{t}$, $b\bar{b}$, $W^+W^-$, and $ZZ$:
 $\left\langle \sigma v
\right\rangle=\sum_{f=t,b,W,Z}\left\langle \sigma v
\right\rangle_{\chi\bar{\chi}\rightarrow f\bar{f}}$. 
The blue region indicates $0.32 < p{\rm -value} < 1.0$ (68.3\% CL), 
the green region $0.05 < p{\rm -value} < 0.317$ (95\% CL), and 
the yellow region $0.01 < p{\rm -value} < 0.05$ (99\% CL).  
The panels are:  upper-left for the combination of NFW-MAX,
upper-right for NFW-MED,
bottom-left for NFW-MIN, and
bottom-right for NFW-DC.  
}
\end{figure}

We keep four annihilation channels ($t\bar{t}$, $b\bar{b}$, $W^+W^-$,
and $ZZ$) of DM for the the calculation of antiproton ratio,
because these channels can generate
antiprotons after decay and hadronization, while the $e^+ e^-$ and
$\mu^+ \mu^-$ channels are irrelevant and the contribution from the
$\tau^+ \tau^-$ channel is negligible. 
We therefore define $\left\langle \sigma v \right\rangle_{\bar{p}/p} \equiv
\sum_{f\bar{f}=t\bar{t},b\bar{b},WW,ZZ}\left\langle \sigma v
\right\rangle_{\chi\bar{\chi}\rightarrow f\bar{f}}$. We take this
multi-annihilation DM scenario to fit to the AMS-02 $\bar{p}/p$ data,
and use only the data with KE larger than 5 GeV in the
fitting. The results in terms of CL regions in the ($M_{\chi}$,
$\left\langle \sigma v \right\rangle$) panels with MAX, MED, MIN, and
DC diffusion models are shown in Fig.~\ref{antip_fit}.

In Fig.~\ref{antip_fit}, we can see a strong and almost linear correlation
between $M_{\chi}$ and $\left\langle \sigma v \right\rangle_{\bar{p}/p}$
in all four panels.
The 95\% CL lower and upper limits on $M_{\chi}$ depends on the diffusion-model.
The 95\% lower limit on $M_{\chi}=$ 850, 600,
1000, and 450 GeV for MAX, MED, MIN, and DC diffusion models, respectively.
The value of $\left\langle \sigma v \right\rangle_{\bar{p}/p}$
required to fit the AMS-02 $\bar{p}/p$ data is always smaller than the
$\left\langle \sigma v \right\rangle_{tot}$ required to fit the AMS-02
$\frac{e^+}{e^++e^-}$ data with the corresponding $M_{\chi}$, except
for the MIN diffusion model with $M_{\chi}\geq$ 10 TeV. 
Nevertheless, the values of
$\left\langle \sigma v \right\rangle_{tot}$ and $\left\langle \sigma v
\right\rangle_{\bar{p}/p}$ are of the same order of magnitude.
The values of $\left\langle \sigma v \right\rangle_{\bar{p}/p}$ for the MIN
diffusion model are about two orders of magnitude larger than the other
models, mainly because of the fact that the diffusion zone of MIN is thinner
than the other diffusion models such that particles are easier to escape
the diffusion zone during propagation. It then requires larger values of
$\left\langle \sigma v \right\rangle_{\bar{p}/p}$ to compensate for the
lost particles.  Thus, for $M_{\chi}\geq 10$ TeV the value of
$\left\langle \sigma v \right\rangle_{\bar{p}/p}$ for the MIN diffusion
model is of the same order as the value of $\left\langle \sigma v
\right\rangle_{tot}$ in Fig.~\ref{nfw_min}.
In other words, in the NFW-MIN case for $M_{\chi}\geq10$ TeV, it
can tolerate large branching ratios of massive vector boson and quark
channels without conflict with the upper bound from AMS-02 $\bar{p}/p$
data.

The 95\% CL upper limit on $\left\langle \sigma v
\right\rangle_{\bar{p}/p}$ shown in Fig.~\ref{antip_fit} are also 
diffusion-model dependent. Taking the MED diffusion model as an example with
$M_{\chi}=$ 1(10) TeV, the upper limit on $\left\langle \sigma v
\right\rangle_{\bar{p}/p}\approx$ 20 (1000) pb, which is about 1/30 of
the favored value of $\left\langle \sigma v \right\rangle_{tot}$ in
the AMS-02 $\frac{e^+}{e^++e^-}$ case. 
In other words, if we try to
fit both AMS-02 $\bar{p}/p$ and $\frac{e^+}{e^++e^-}$ data, the
$\textrm{BR}({\bar{q}q})+\textrm{BR}({VV}) \lesssim 3\%$ or ${\rm
  Br}(l^+l^-)\gtrsim 97\%$ is required. However the panel of (${\rm
  Br}(l^+l^-)$, p) in Fig.~\ref{nfw_med} tells us that the parameter region
satisfying the above requirement is very narrow: only when
$M_{\chi}\lesssim$ 1 TeV and the $p-$value is less than 0.1. 
After scrutinizing the parameter space of the multi-annihilation channel DM
scenario, it cannot provide simultaneously
good fits to both the AMS-02 $\frac{e^+}{e^++e^-}$
and $\bar{p}/p$ data with a $p-$value larger than 0.05 
based on the NFW-MED combination.

A simultaneous solution for both AMS-02 $\frac{e^+}{e^++e^-}$ and 
$\bar{p}/p$ data could be found under the NFW-MIN combination for
$M_{\chi}\subset [10,30]$ TeV, as shown by the compensating color 
in the panel of
($M_{\chi}$, $\left\langle \sigma v \right\rangle$) in
Fig.~\ref{nfw_min}. The upper two panels in
Fig.~\ref{both_positron_antip} show the a benchmark point of the
solution for both AMS-02 $\frac{e^+}{e^++e^-}$ and $\bar{p}/p$ data,
with both $p$-values larger than 0.05: $M_{\chi}=$ 23 TeV with
composition of $22\%\,\tau^+\tau^-$, $39\%\,t\bar{t}$, and $39\%\,ZZ$
channels.

After the comprehensive parameter survey of the multi-annihilation channel
DM, we did not find any solution to simultaneously explain the AMS-02
$\frac{e^+}{e^++e^-}$ and $\bar{p}/p$ data under the case of the
NFW profile combined with MAX, MED, or DC diffusion models.

\section{Different Dark Matter Density profiles}

So far in the above sections, we only focus on the NFW DM
density profile. In this section, we 
investigate the consequences of using different density profiles by
adopting the isothermal and Einasto profiles.

Among these profiles, we have normalized the DM density
to $\rho(r_\odot=8.5\, \rm{kpc})\,=\,0.3\, \rm{GeV cm^{-3}}$
at our solar system, where $r_\odot$ is the distance of our solar
system from the GC.
Note that the largest uncertainty among
various profiles happens within 1 kpc around the GC.
Around the vicinity of our solar system, 
the isothermal and NFW profiles are almost identical, but 
however some deviations between Einasto and NFW happens. 
Hence, the positron spectra from DM annihilation from
isothermal and NFW profiles are quite similar, but different from
the Einasto profile. In general, the Einasto profile will give a 
softer positron spectrum than the NFW and isothermal. 
This feature could make it easier for the multi-annihilation 
channel DM scenario to fit simultaneously to both 
AMS-02 $\frac{e^+}{e^++e^-}$ and $\bar{p}/p$ data, 
because the missing ingredient in the NFW profile is that 
the positron spectrum from leptonic channels is not soft enough for the
$\frac{e^+}{e^++e^-}$ data. 
In the following, we take the
combination of the Einasto profile and the DC diffusion model to demonstrate
this feature, since the DC model utilizes a more recent B/C
observation to constrain their cosmic-ray propagation parameters.

\begin{figure}[th!]
\centering
\includegraphics[width=3.2in]{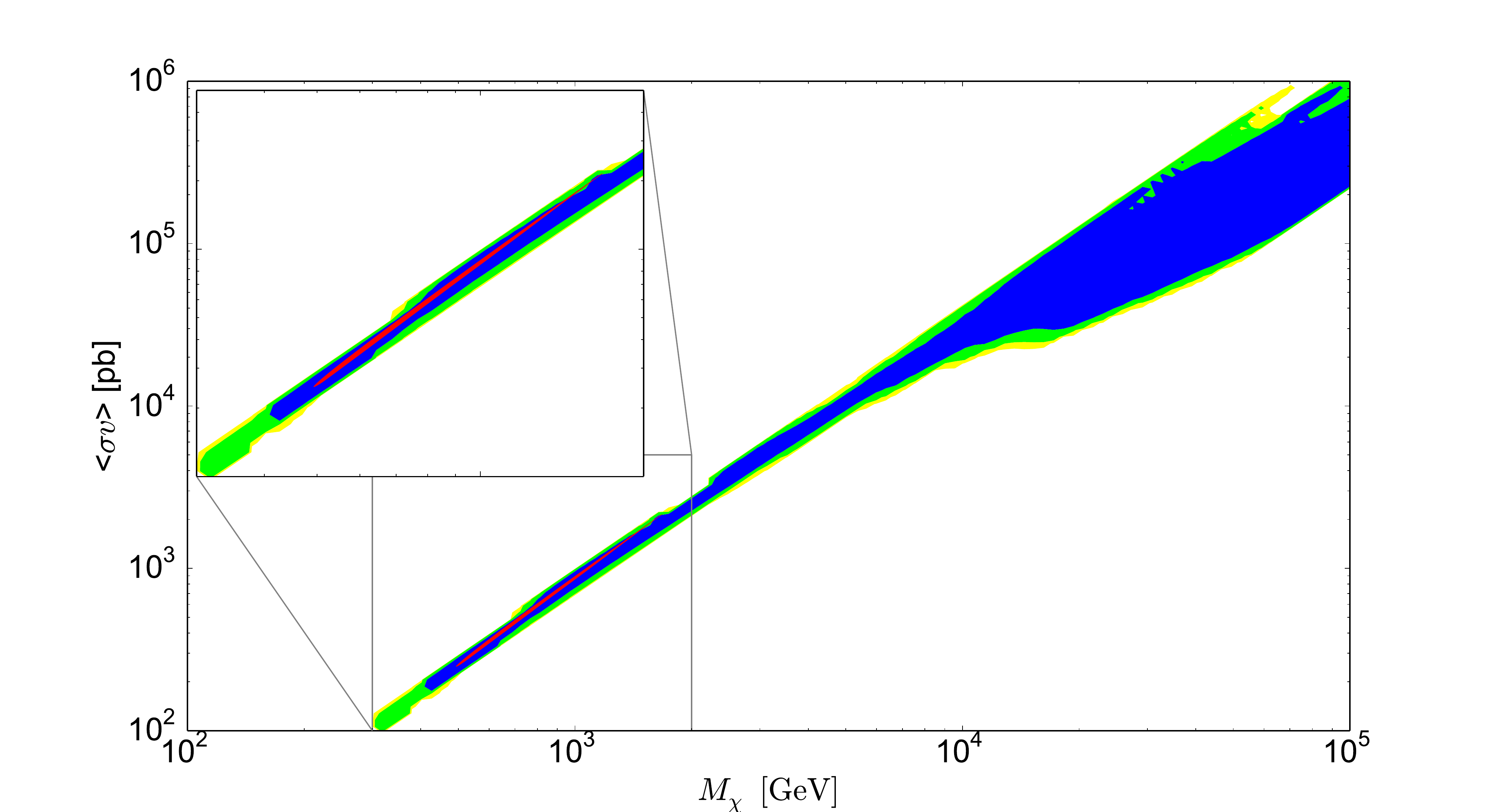}
\includegraphics[width=3.2in]{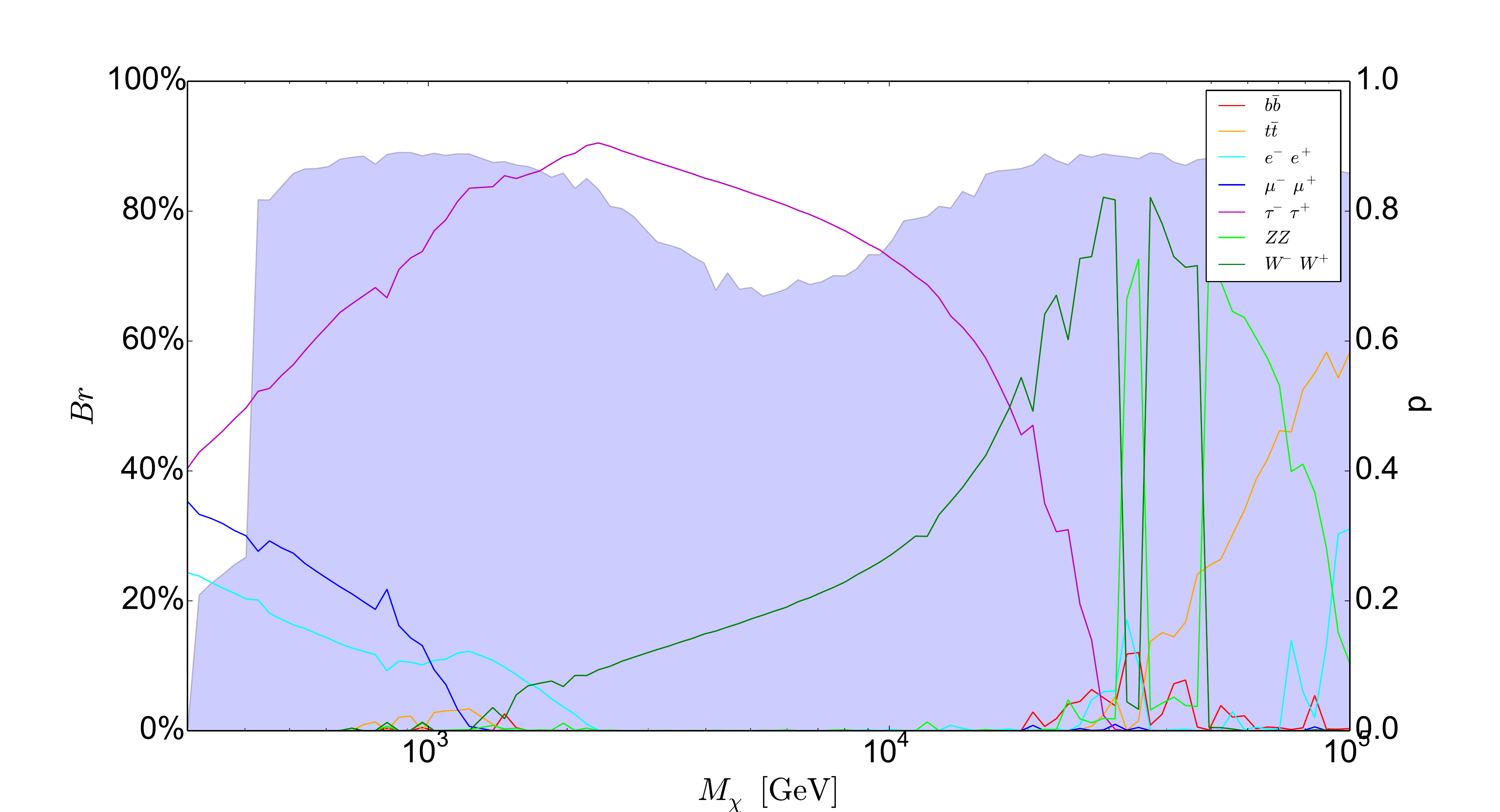}
\includegraphics[width=3.2in]{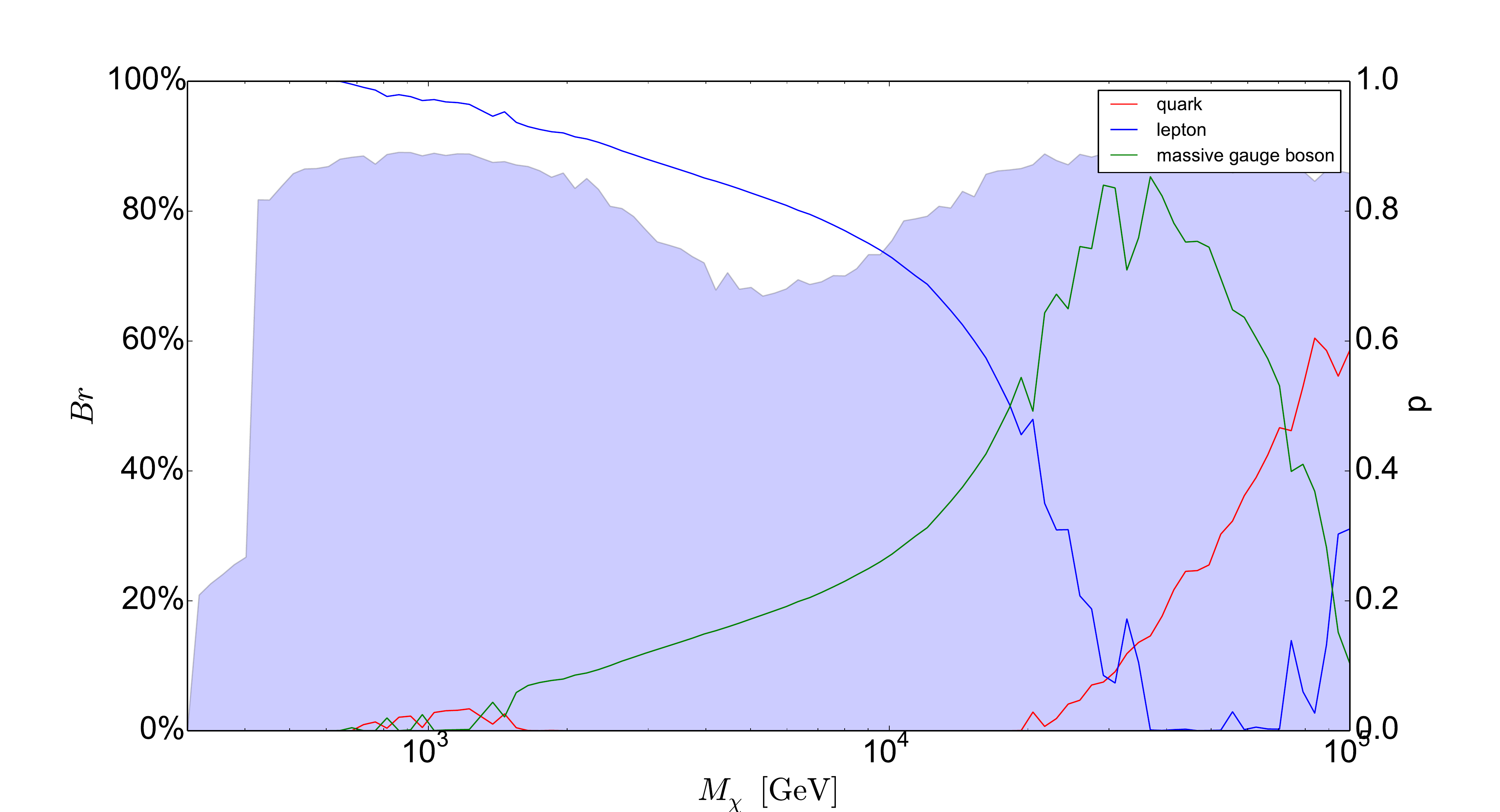}
\includegraphics[width=3.2in]{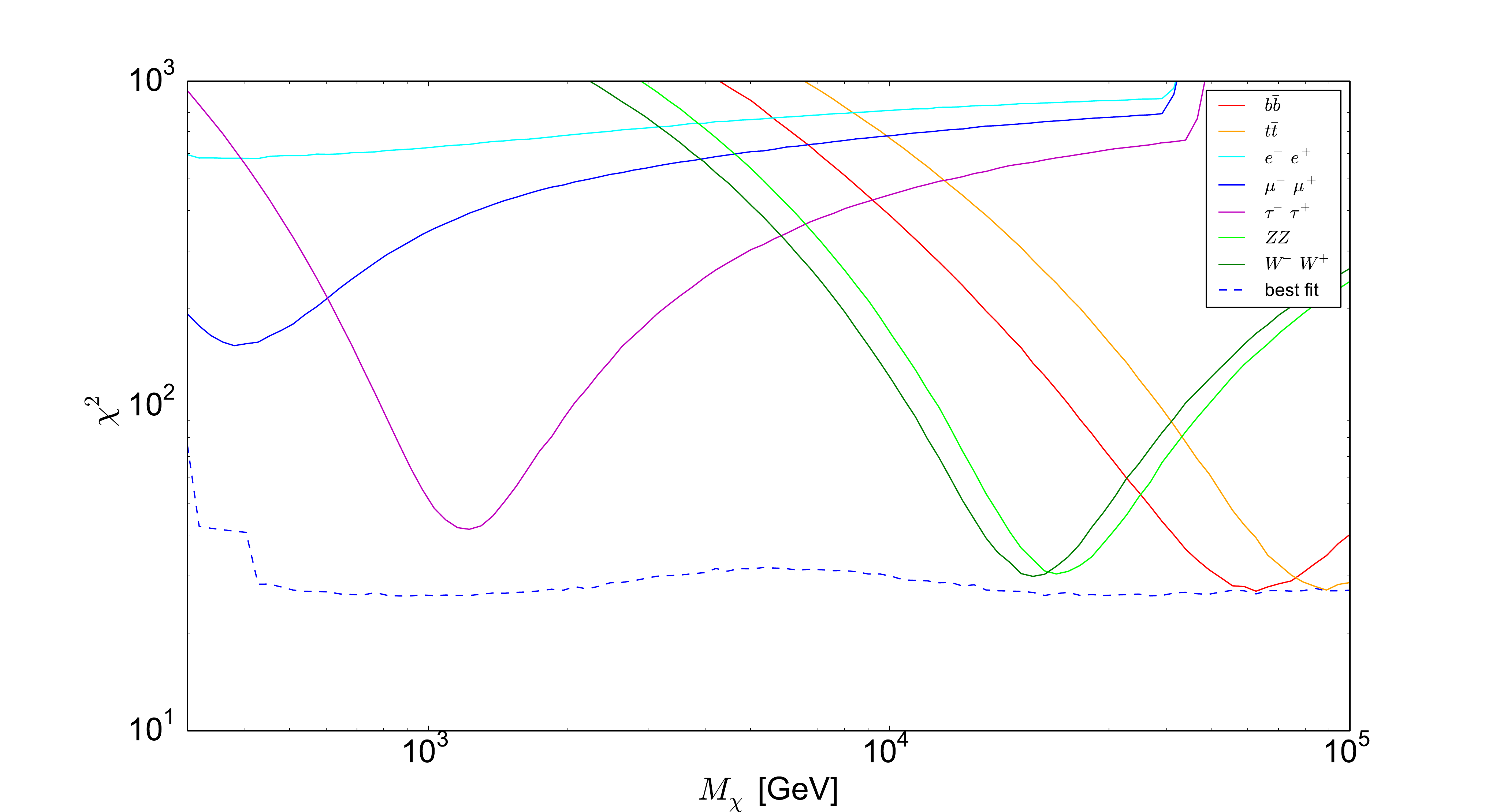}
\caption{ \small \label{ein_dc_positron} 
The same as Fig.~\ref{nfw_med} but for the combination of of Einasto-DC, except
without the last panel.
The red region in the upper-left panel in the plane of 
($M_{\chi}$, $\left\langle \sigma v \right\rangle$)
shows the allowed region of simultaneously fitting to both AMS-02
$\frac{e^+}{e^-+e^+}$ and $\bar{p}/p$ datasets with $p$-value $>$ 0.05
for each dataset.  A zoom is displaced in the upper-left corner of
that panel.
}
\end{figure}

\begin{figure}[th!]
\centering
\includegraphics[width=3.2in]{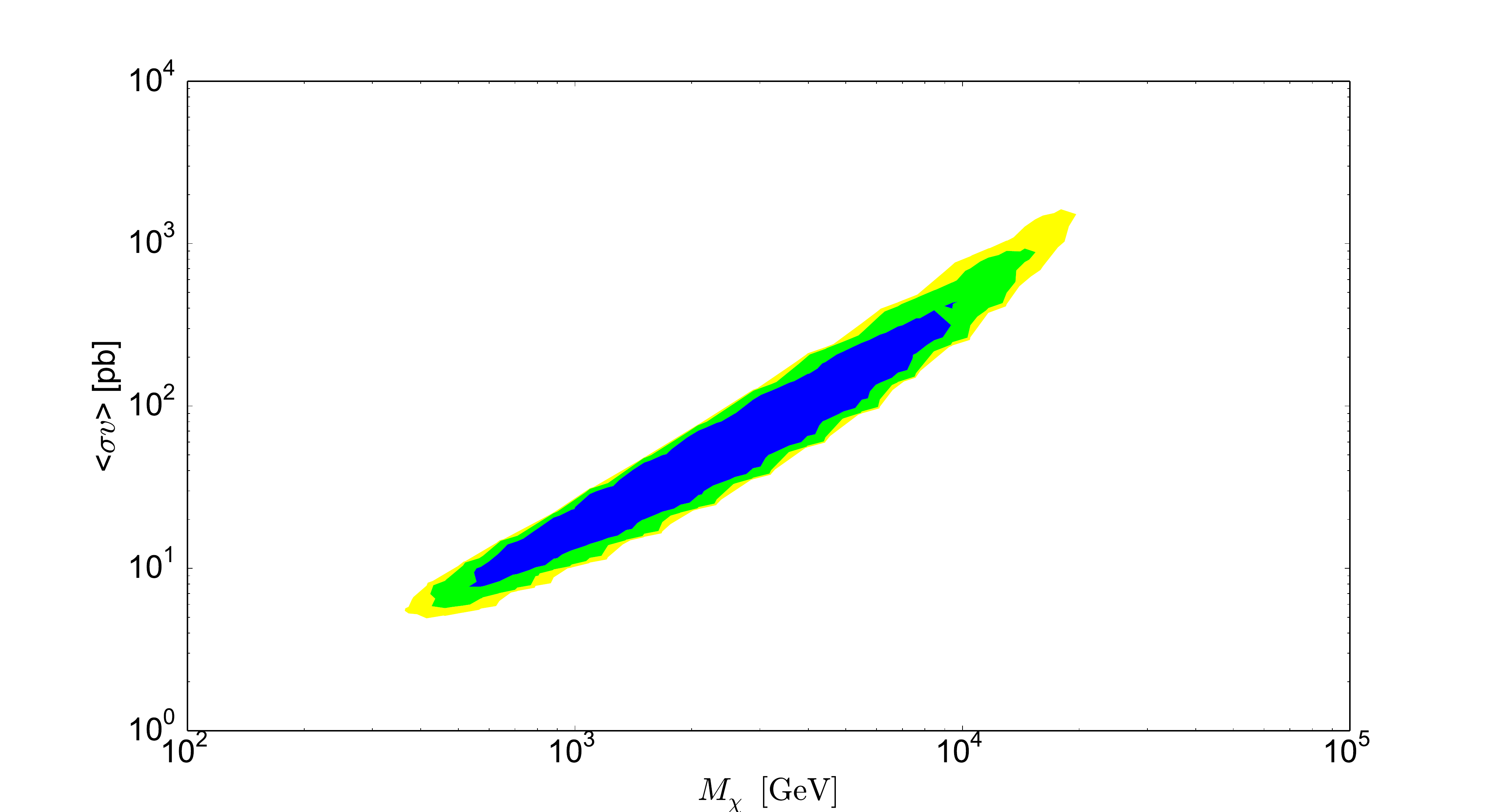}
\includegraphics[width=3.2in]{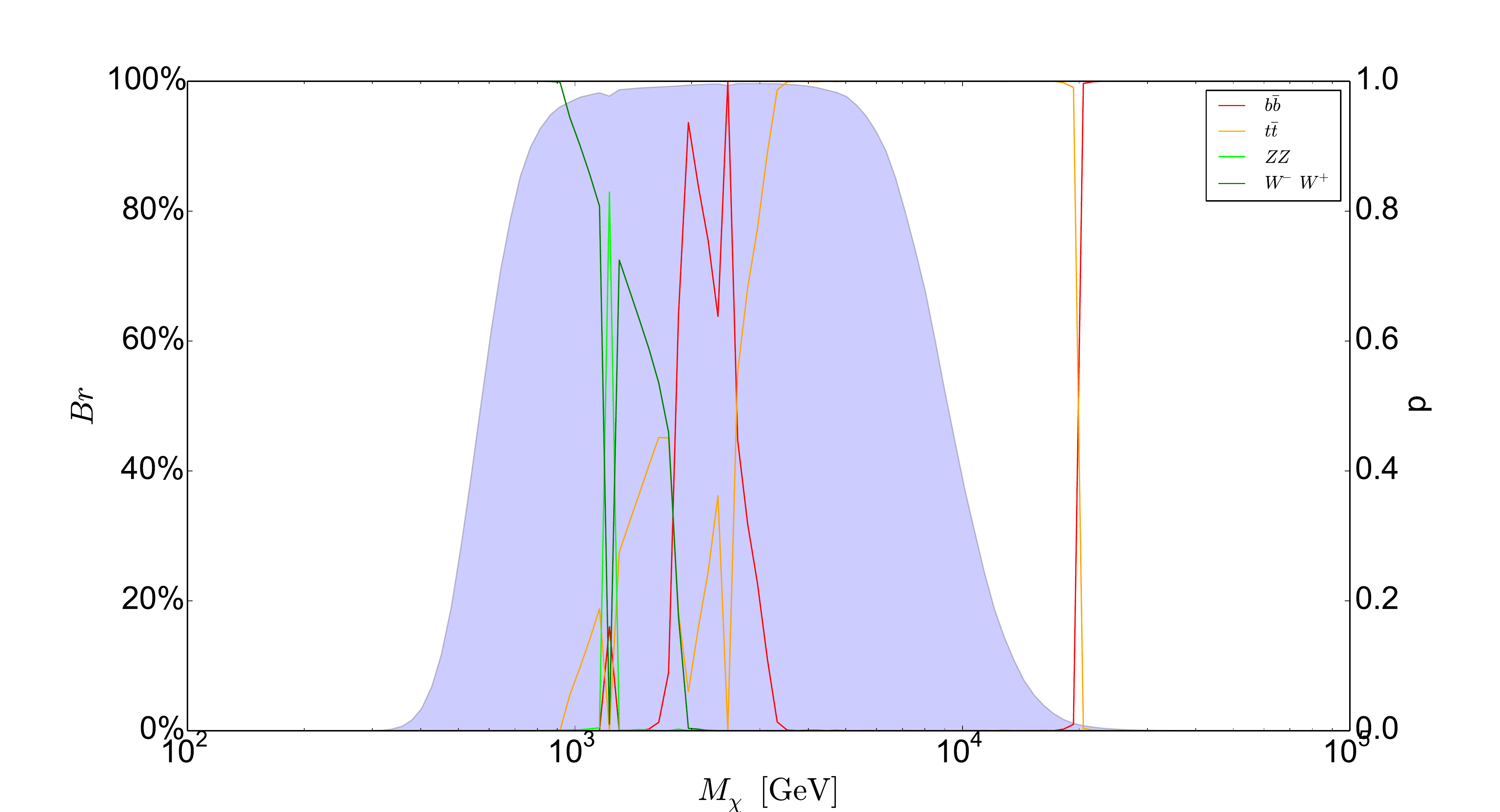}
\includegraphics[width=3.2in]{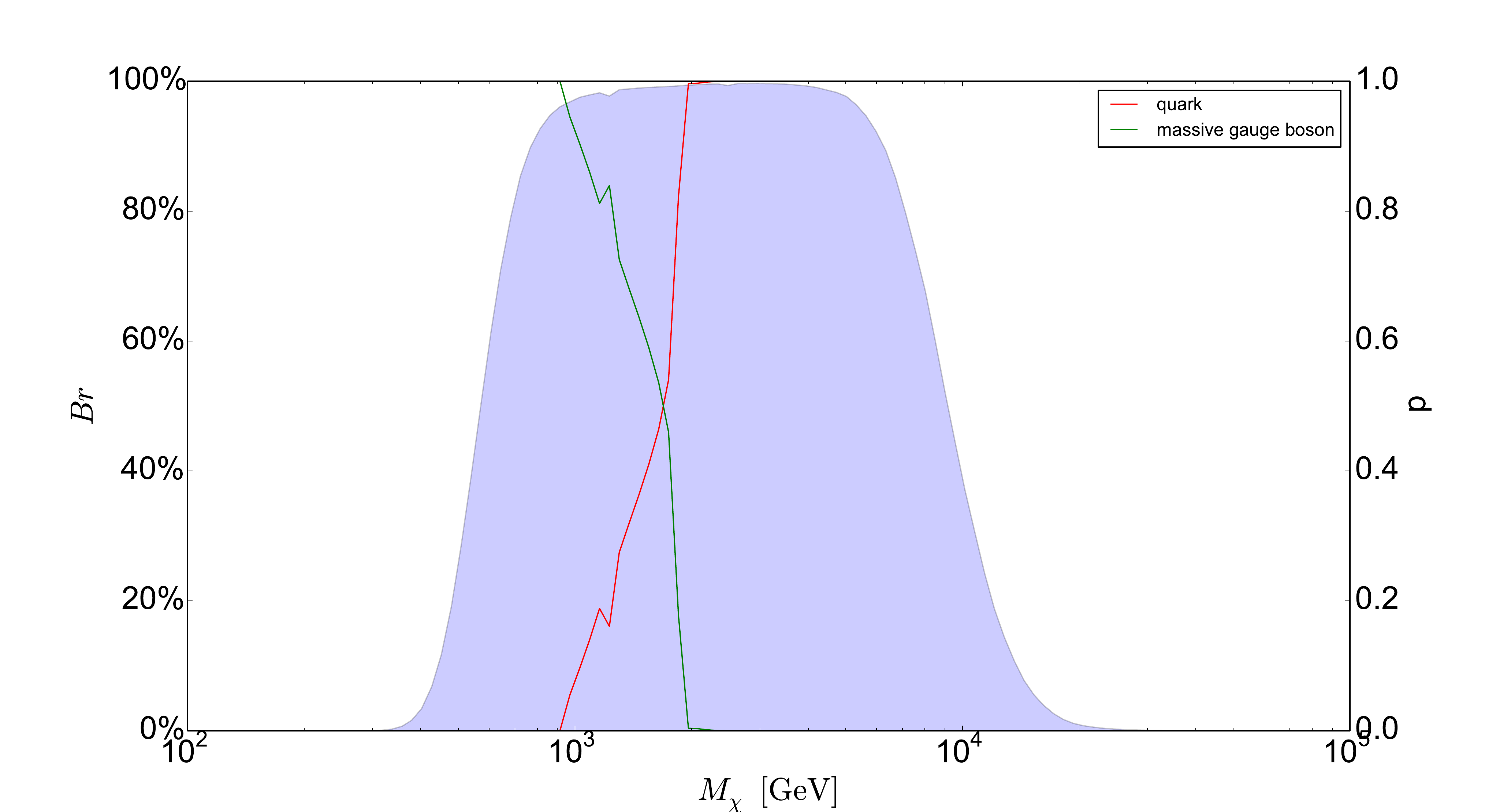}
\caption{ \small \label{ein_dc_antip} 
Fitting results using the combination of Einasto-DC with respect to the
AMS-02 $\bar{p}/p$ data.
Upper-left: ($M_{\chi}$, $\left\langle \sigma v \right\rangle$) -- 
confidence-level regions, labeled in the same way as Fig.~\ref{antip_fit}. 
Upper-right: ($M_{\chi}$, Br) -- fitting to the AMS-02 $\bar{p}/p$ data
alone with $t\bar{t}$, $b\bar{b}$, $W^+W^-$, and $ZZ$ channels.
Lower: ($M_{\chi}$, Br) -- similar to the upper-right one, but 
summing over the similar channels: quark = $\sum_{f=b,t}{\rm Br}(f\bar{f})$
and massive gauge boson = $\sum_{f=W,Z}{\rm Br}(f\bar{f})$.
}
\end{figure}

\begin{figure}[th!]
\centering
\includegraphics[width=3.2in]{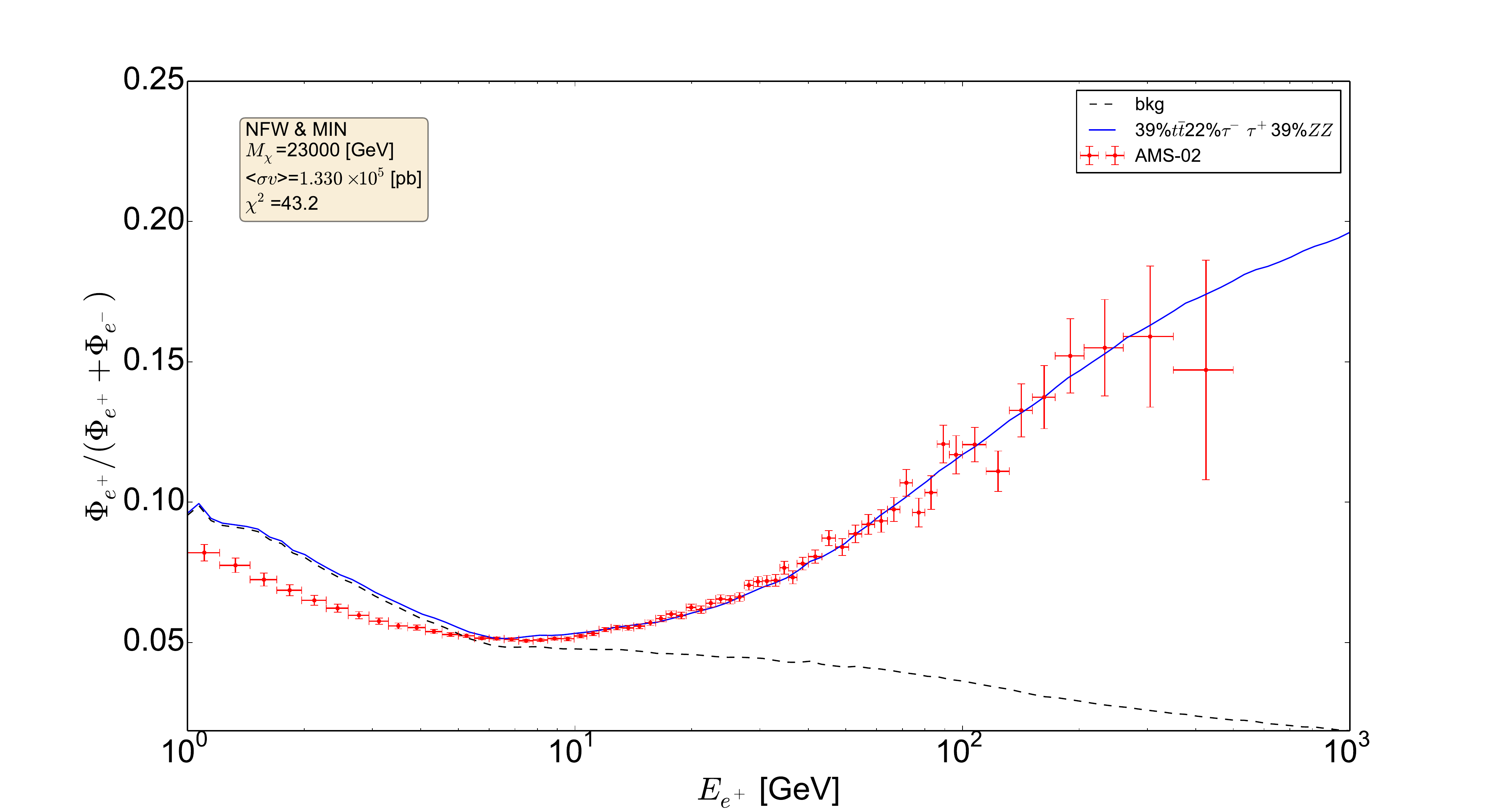}
\includegraphics[width=3.2in]{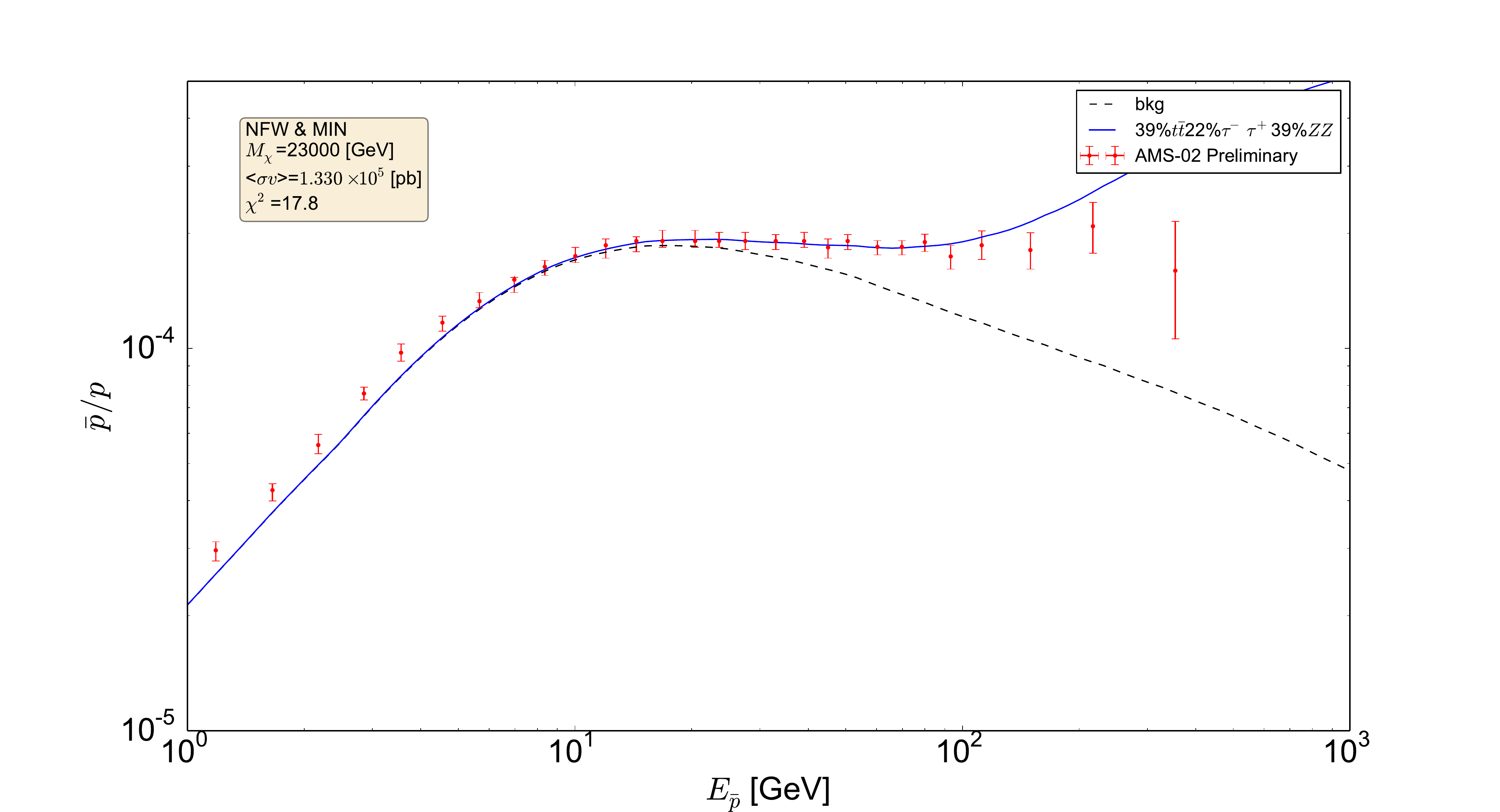}
\includegraphics[width=3.2in]{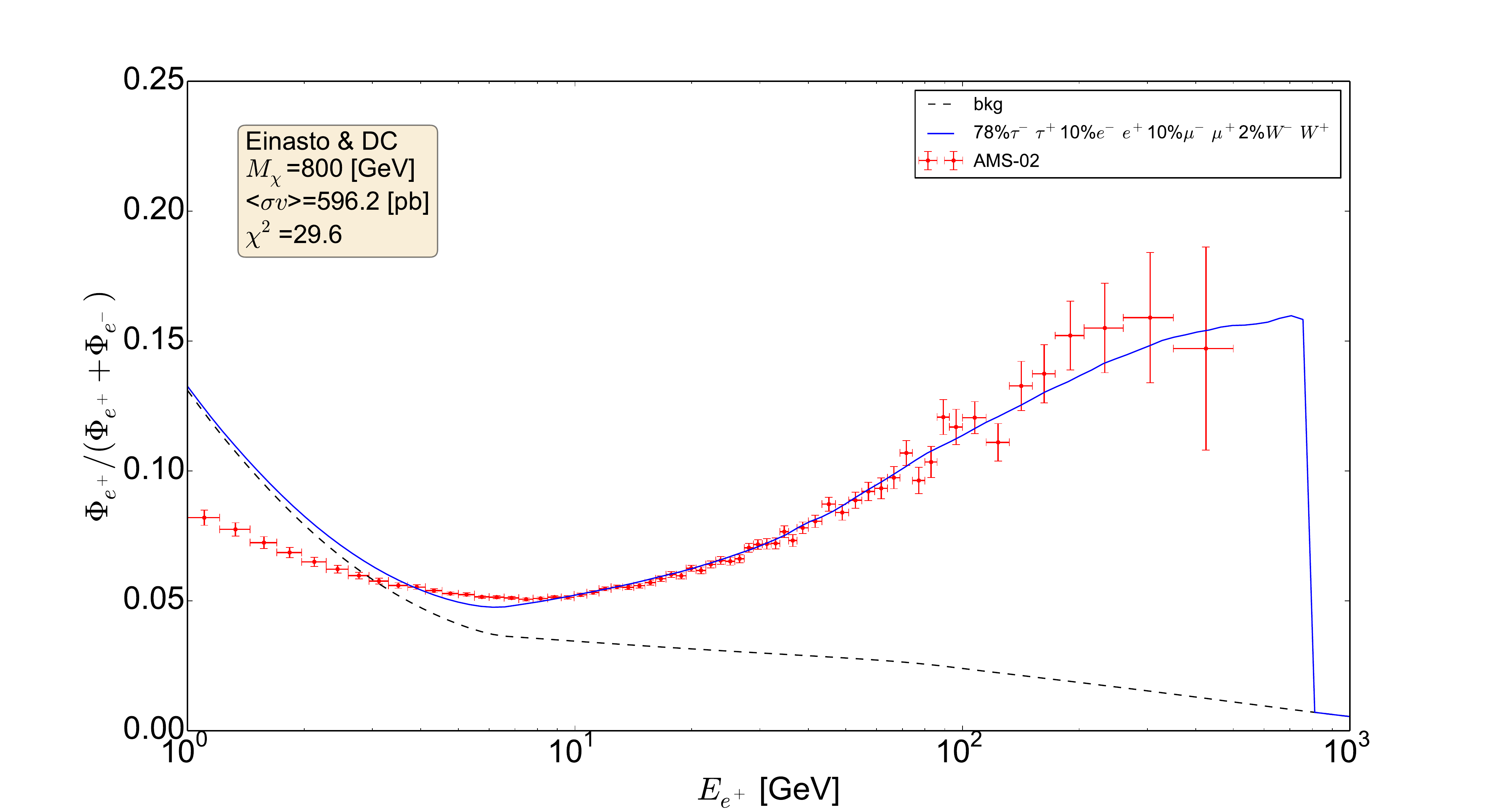}
\includegraphics[width=3.2in]{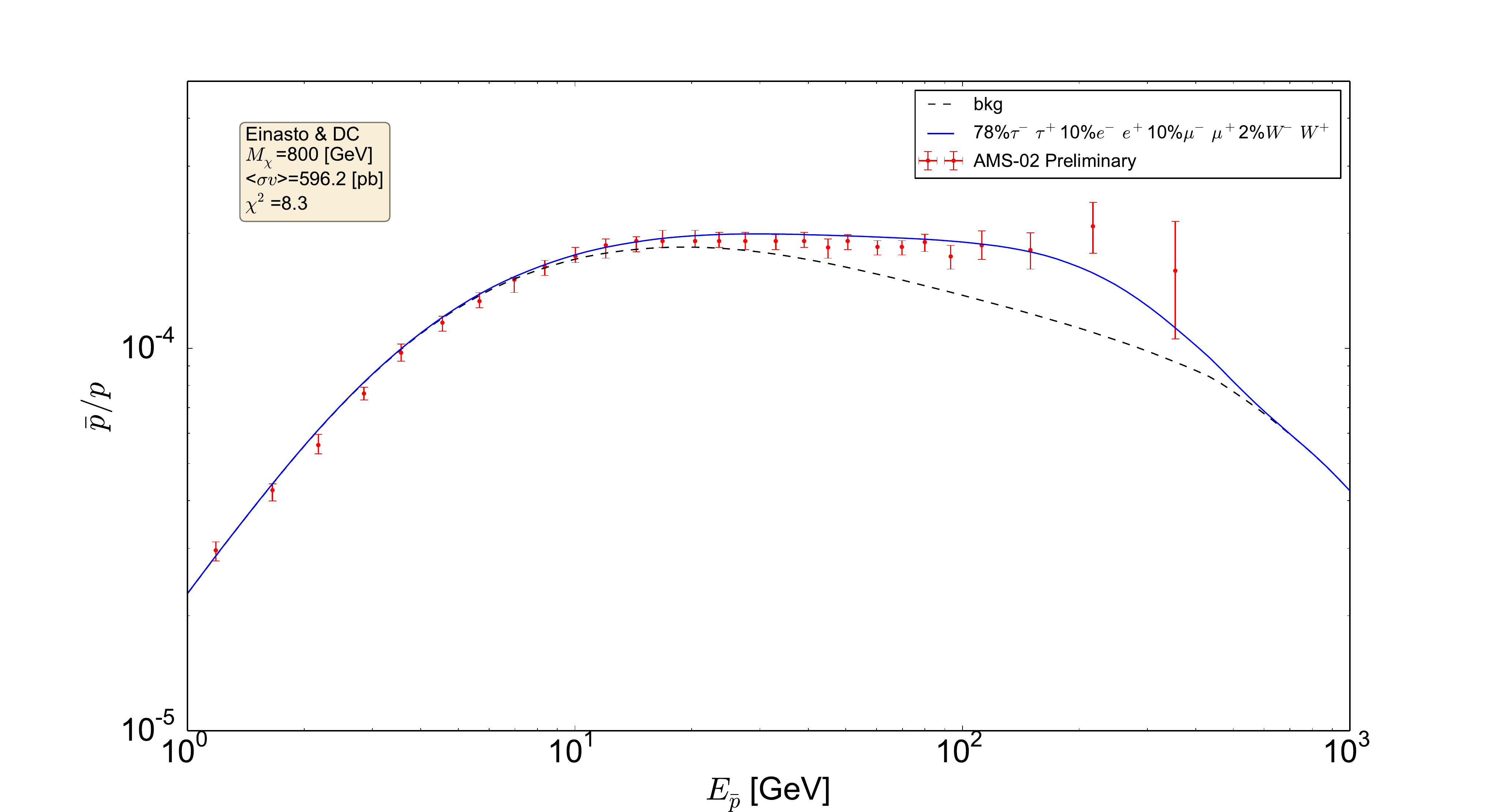}
\caption{ \small \label{both_positron_antip} 
The left-handed panels are for the fitting results to 
the AMS-02 $\frac{e^+}{e^++e^-}$ data, while 
the right-handed panels are for the fitting results to
the AMS-02 $\bar{p}/p$ data.  
Upper two panels: under the NFW-MIN combination with a benchmark point
within the compensating color region in the ($M_{\chi}$,
  $\left\langle \sigma v \right\rangle$) panel in Fig.~\ref{nfw_min}
$\biggr [  M_{\chi}=23$ TeV, $\left\langle \sigma v \right\rangle_{tot}=1.330
    \times 10^5$ pb, ${\rm Br}(\tau^+\tau^-)=22\%$, ${\rm
      Br}(ZZ)=39\%$, ${\rm Br}(\bar{t}t)=39\% \biggr ]$.  
Lower two panels: under the Einasto-MED combination with a benchmark point
within the red contour in the ($M_{\chi}$, $\left\langle \sigma v
  \right\rangle$) panel in Fig.~\ref{ein_dc_positron}
 $\biggr [ M_{\chi}=800$ GeV, $\left\langle \sigma v
                 \right\rangle_{tot}=5.962 \times 10^2$ pb, ${\rm
                   Br}(e^+e^-)=10\%$, ${\rm Br}(\mu^+\mu^-)=10\%$,
                 ${\rm Br}(\tau^+\tau^-)=78\%$, ${\rm
                   Br}(W^+W^-)=2\% \biggr ]$.  
}
\end{figure}

Based on the Einasto-DC combination, the AMS-02 $\frac{e^+}{e^++e^-}$
and $\bar{p}/p$ fitting results are shown in Figs.~\ref{ein_dc_positron} and
\ref{ein_dc_antip}, respectively.
 The benchmark point of
multi-annihilation channel DM that can explain both data sets is illustrated
in the lower two panels of Fig.~\ref{both_positron_antip}.
Comparing the Fig.~\ref{nfw_dc} (NFW) with Fig.~\ref{ein_dc_positron} 
(Einasto), larger values of $M_{\chi}>$ 10 TeV are allowed at 95\% CL 
in the Einasto profile. From the panels of ($M_{\chi}$, $\chi^2$), 
the fitting of the
single channels of $\tau^+\tau^-$, $VV$ and $\bar{q}q$ are all
improved in the Einasto profile than those in NFW one. Also, because of the
improvement in $VV$ and $\bar{q}q$ channels, the region of $M_{\chi}>$
10 TeV is allowed.
Another difference in the panels of ($M_{\chi}$, $\chi^2$) between
these two profiles is that the corresponding values of $M_{\chi}$ of the
minimum of each channel shifts to larger values in the Einasto
case. This comes from the fact that the Einasto profile makes positron spectra
from each channels softer.
In the panels of ($M_{\chi}$, Br) in Fig.~\ref{ein_dc_positron}, when
$M_{\chi}$ less than 1200 GeV, the ${\rm Br}(l^+l^-)$ of the best-fit
point can be larger than 95\%, and so  ${\rm Br}(VV)+{\rm
  Br}(\bar{q}q)$ is suppressed below 5\%. This region provides a  nice
habitat to accommodate the AMS-02 $\bar{p}/p$ data. In the panel of
($M_{\chi}$, $\left\langle \sigma v \right\rangle$), the red
long-elliptical contour indicates the allowed DM mass and annihilation
cross section that can explain well these two AMS-02 observations. The DM
mass range is $M_{\chi}\subset[500,1500]$ GeV.

In Fig.~\ref{ein_dc_antip}, we show the results of fitting to the AMS-02
$\bar{p}/p$ data under the Einasto-DC combination. Comparing the panel
of ($M_{\chi}$, $\left\langle \sigma v \right\rangle$) in this figure
to the bottom-right panel 
(corresponding to the NFW-DC combination) in Fig.~\ref{antip_fit},
the 95\% CL allowed range for DM mass does not
change much, 400 GeV $\lesssim M_{\chi}\lesssim$ 15 TeV. However,
because of the DM density $\rho(r)$ of the Einasto profile is larger than
that of NFW for $r\leq r_{\odot}$, the entire CL region shifts down
by a factor of 2 for the Einasto case, indicating that the value of
$\left\langle \sigma v \right\rangle_{tot} $ for the Einasto case is half
of that for NFW. 
Similar observation can be seen from the fitting
to the AMS-02 $\frac{e^+}{e^++e^-}$ data, by comparing the panels of
($M_{\chi}$, $\left\langle \sigma v \right\rangle$) in
Figs.~\ref{nfw_dc} and ~\ref{ein_dc_positron}.
Another observation from the two panels of ($M_{\chi}$, Br) in
Fig.~\ref{ein_dc_antip} for the best-fit point is that when
$M_{\chi}\lesssim$ 1700 GeV, the massive gauge boson channels take the
most responsibility to fit the data.  However, when $M_{\chi}\gtrsim$ 1700 GeV,
the quark channels take over the major role to do the
fitting. 

In summary, the two major differences between the Einasto and NFW profiles
are (i) the former makes the positron spectra from DM annihilation
softer than the latter, and (ii) in order to produce the same amount of
positron or antiproton flux from DM annihilation, a slightly smaller
annihilation cross section is required in the Einasto profile.

\section{Conclusions}

In this work, we have postulated a DM scenario with multi-annihilation
channels to account for the positron fraction
excess and antiproton-to-proton ratio observed in cosmic rays. By
using the most updated $\frac{e^+}{e^++e^-}$ and
$\bar{p}/p$ data from AMS-02, we have performed a comprehensive analysis to
determine the preferred DM annihilation channel in terms of statistic
quantities $\chi^2$ and $p$-values.
We have considered the DM annihilation into $t\bar{t}$, $b\bar{b}$, $W^+W^-$,
$ZZ$, $e^+e^-$, $\mu^+\mu^-$, and $\tau^+\tau^-$ channels from a single
annihilation process, and taken the branching ratio of each channel and
the total cross section as the fitting parameters.
We focus on two aspects: i) Which channel is preferred by the AMS-02
$\frac{e^+}{e^++e^-}$ data, and how strong the confinement of
branching ratios can be implied. ii) Under this multi-annihilation
channel DM scenario, we are looking for a simultaneous solution for 
both AMS-02 $\frac{e^+}{e^++e^-}$ data and $\bar{p}/p$ data.  
Several diffusion models (MAX, MED, MIN, and DC) and DM
density profiles (isothermal, NFW, Einasto) are invstigated in our
analysis. 
Uncertainties in astrophysical backgrounds are obtained 
via combinations of various density profiles and diffusion models.

Our findings are summarized as follows.
\begin{enumerate}
\item 
The allowed region of DM mass $M_{\chi}$ and annihilation cross
sections $\left\langle \sigma v \right\rangle_{tot}$ that can provide
good fits with $p$-value $\geq$ 0.05 to the AMS-02 $\frac{e^+}{e^++e^-}$
data is very broad. The range of $M_{\chi}$ extends from 300 GeV to a 
few tens of TeV, and $\left\langle \sigma v \right\rangle_{tot}$ extends 
from $O(100)$ pb to $O(10^5)$ pb. 
%
However, there is strong and almost linear correlation between the
$M_{\chi}$ and $\left\langle \sigma v \right\rangle_{tot}$.

\item
The DM mass below 300 GeV is ruled out at 95\% CL by the
$\frac{e^+}{e^++e^-}$ data, and this lower limit of $M_{\chi}$ is
diffusion-model and density-profile independent. On the other hand,
the upper limit for the DM mass strongly depends on the diffusion
models and density profiles.
Pure leptonic channels are solutions of AMS-02 positron fraction data 
under combination of NFW-MIN with $M_{\chi}\subset$ [300, 450] GeV or 
combination of Einasto-DC with $M_{\chi}\subset$ [300, 1800] GeV.

\item
The high statistical measurement from AMS-02 has the capacity to
distinguish among different annihilation channels. For a few hundred
GeV to several TeV DM, the $\tau+\tau^-$ channel is the ideal one to
fit to the $\frac{e^+}{e^++e^-}$ data. On the other hand, for several
TeV to 10 TeV DM, the $W^+W^-$ and $ZZ$ are the favorites. Beyond 10 TeV
the $b\bar{b}$ and $t\bar{t}$ channels begin to take over
and completely dominate very quickly.

\item
Based on the multi-annihilation channel DM scenario, for all MAX, MED,
and MIN diffusion models, when $M_{\chi}\simeq$ 1 TeV, the AMS-02
$\frac{e^+}{e^++e^-}$ data has the best confinement of the branching
ratio of the sum of three leptonic channels, and ${\rm Br}(l^+l^-)$ is
within $\pm 20\%$ deviation from the best-fit value at 95\%
CL. However, the value of ${\rm Br}(l^+l^-)$ for the best-fit point
depends on the diffusion model.

\item
The DM scenario with multi annihilation channels under the
combinations of the MAX, MED and DC diffusion models, and the NFW
density profile cannot satisfy both AMS-02 $\frac{e^+}{e^++e^-}$ and
$\bar{p}/p$ data at 95\% CL.  However, under the NFW-MIN
and Einasto-DC combinations, we find windows in, respectively,
$M_{\chi}\subset[10,30]$ TeV and $M_{\chi}\subset[500,1500]$ GeV
that can fit well with $p$-value $\geq$ 0.05 to both datasets,
For the Einasto-DC case, the $\tau^+\tau^-$ channel dominates and the
massive gauge boson and quark channels are suppressed. On the other hand,
for the NFW-MIN case the $ZZ$ and $t\bar{t}$ are the two major
channels.

\end{enumerate}

During the last stage of the manuscript, we came across a few similar works
in model-independent approach
\cite{ams02_antip_1,new1,new2,new3} and in specific models \cite{new4,new5}.

\section*{Acknowledgment}  
This work was supported by the MoST
of Taiwan under Grants No. 102-2112-M-007-015-MY3.


\end{document}